\documentclass[aps,pra,twocolumn,superscriptaddress,amsmath,amssymb,]{revtex4}

\usepackage{graphicx}
\usepackage{epstopdf}
\usepackage{lmodern}
\usepackage{xfrac}
\usepackage{color}

\newcommand{\OD}[0]{\mathrm{OD}}
\newcommand{\appropto}{\mathrel{\vcenter{
  \offinterlineskip\halign{\hfil$##$\cr
    \propto\cr\noalign{\kern2pt}\sim\cr\noalign{\kern-2pt}}}}}

\usepackage{natbib}

\usepackage{hyperref}

\begin{document}
\title{Characterization of a photon pair source based on a cold atomic ensemble using a cascade level scheme}

\author{Alessandro~Cer\`{e}}
\affiliation{Centre for Quantum Technologies, National University of Singapore, 3 Science Drive 2, Singapore 117543}
\author{Bharath~Srivathsan}
\affiliation{Centre for Quantum Technologies, National University of
  Singapore, 3 Science Drive 2, Singapore 117543}
\affiliation{current address: Max Planck Institute for the Science of Light, 91058 Erlangen, Germany}
\author{Gurpreet~Kaur~Gulati}
\affiliation{Centre for Quantum Technologies, National University of
  Singapore, 3 Science Drive 2, Singapore 117543}
\affiliation{current address: Jet Propulsion Laboratory, Caltech, Pasadena, California 91109, USA}
\author{Brenda~Chng}
\affiliation{Centre for Quantum Technologies, National University of Singapore, 3 Science Drive 2, Singapore 117543}
\author{Christian~Kurtsiefer}
\affiliation{Centre for Quantum Technologies, National University of Singapore, 3 Science Drive 2, Singapore 117543}
\affiliation{Department of Physics, National University of Singapore, 2 Science Drive 3, Singapore 117551}

\email[]{christian.kurtsiefer@gmail.com}
\date{\today}
\begin{abstract}
  We characterize a source of photon pairs based on cascade decay in a cold $^{87}$Rb ensemble.
  This source is particularly suited to generate of photons for
  interaction with $^{87}$Rb based atomic systems.
  We experimentally investigate the dependence of pair generation rate, single
  photon heralding efficiency, and bandwidth as a function of the number of
  atoms, detuning and intensity of the pump beams.
  The observed power and detuning behaviors can be explained by
  the steady state solution of an established three-level model of an atom.
  Measurements presented here
  provide a useful insight on the optimization of
  this kind of photon pair sources.
\end{abstract}


\maketitle

\section{Introduction}

  Time-correlated and entangled photon pairs are an important resource for a wide
  range of quantum optics experiments, ranging from fundamental tests~\cite{Clauser:1978jf,Aspect:1981ga}
  to applications in quantum information~\cite{Ekert:1991zz,Bouwmeester:1997wk,Boschi:1998iu}.
  A common method to obtain photon pairs is
  Spontaneous Parametric Down Conversion (SPDC) in
  nonlinear optical crystals~\cite{Burnham:1970gz},
  which have proven to be extremely useful. However, photons prepared by SPDC
  typically have spectral bandwidths ranging from
  0.1~THz to 2~THz~\cite{Kwiat:1995ub,Kurtsiefer:2001ha},
  making interaction with atomic systems with a lifetime-limited bandwidth on
  the order of few MHz difficult.
  Possible solutions to match the bandwidth requirements include the use of
  optical cavities around the
  crystal~\cite{Kuklewicz:2006gq,Wolfgramm:08,Fekete:2013kr},
  filters~\cite{NeergaardNielsen:2007wa,Haase:2009ez},
  and recently the use of miniature monolithic resonators made of nonlinear
  optical materials~\cite{Schunk:2016fl}.
A different approach uses directly atomic systems as the non-linear optical medium
in the parametric process. There, a chain of near-resonant
optical transitions provides an optical nonlinearity that has long been used
for frequency mixing in otherwise inaccessible spectral domains. When two of
the participating modes are not driven, such systems can be used for
photon pair generation via a parametric conversion
process~\cite{Braje:2004dt,Matsukevich:2005fz,Chen:2012jv}.
  As the effective nonlinearity decays quickly with the
detuning from an atomic transition, the resulting photon pairs can be
spectrally very narrow.

  In this work, we investigate such a photon pair source based on four-wave
  mixing in a cold atomic ensemble. The resulting photon pairs are therefore
  directly compatible with ground state transitions of $^{87}$Rb, and the pair
  preparation process does not suffer any reduction in brightness caused by
  additional filtering. This can be interesting for preparing photon states
  that are fragile with respect to linear losses.
  A basic description of the source is presented in~\cite{Srivathsan:2013fa}.

 This source has already been used, with minor modifications, to obtain heralded single photons with an exponentially rising time envelope~\cite{GK_2014,Srivathsan:2014jx}. We have also studied the amount of polarization entanglement in the generated photon pairs, and observed quantum beats between possible decay paths~\cite{Gulati:2015ee}.
  The same source has also been used in conjunction with a separate atomic system, a single $^{87}$Rb atom trapped in a far off resonant focused beam to study their compatibility~\cite{Leong:2015eb} and the dynamics of the absorption of single photons by an atom~\cite{Leong:2016jb}.
  There, we explored a
  limited range of experimental parameters, optimized to observe the physical properties of the biphoton state of interest.
  In this article we present
  a systematic characterization of the source as function of the accessible experimental parameters.
  We believe that
  our scheme is a useful tool for the studies of the interaction of single photons with single and ensembles of atoms.
  In order to characterize the source, we focus our attention on generation rate, heralding efficiency, and the compromise between rates and bandwidth.

  We start with a brief review of the photon pair generation process,
  followed by a presentation of
  the experimental setup, highlighting some of its relevant and differentiating features,
  and a description of the measurement technique.
  The rest of the paper covers systematic variations of the source
  parameters, and their impact on the rates and bandwidth of the emitted
  photon pairs.

\section{Four wave mixing in cold~\texorpdfstring{$^{87}$R\lowercase{b}}{87Rb} based on cascade decay}\label{sec:intro}
  \begin{figure}
    \centering
    \includegraphics[width=\columnwidth]{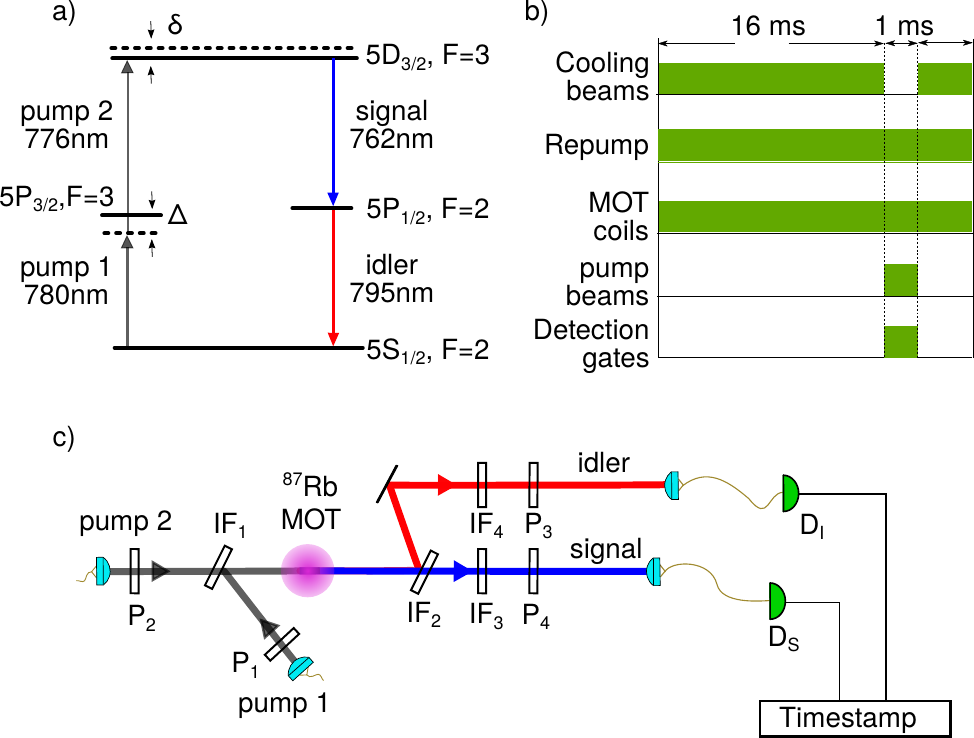}
    \caption{\label{fig:setup}
      (a)~Cascade level scheme used for parametric conversion in atoms.
      (b)~Timing sequence of the experiment.
      (c)~Schematic of the experimental set up, with~P1, P2, P3, and~P4: Polarization filters,~
      IF$_1$, IF$_2$, IF$_3$, and~IF$_4$: interference filters,
      ~D$_{\text{I}}$, D$_{\text{S}}$: avalanche photodetectors.}
  \end{figure}

  The photon pair source in this work is based on the~$\chi^{(3)} $ non-linear susceptibility of~$^{87}$Rb.
  A similar scheme was initially demonstrated with a different choice of transitions and, consequently, wavelengths~\cite{Chaneliere:2006}.
  The relevant electronic structure
  is shown in Fig.~\ref{fig:setup}(a).
  Two pump beams of wavelength 780\,nm (pump~1) and 776\,nm (pump~2) excite the atoms from~$5\text{S}_{\sfrac{1}{2}},F\!=\!2$ to~$5\text{D}_{\sfrac{3}{2}},F\!=\!3$ via a two-photon transition.
  The 780\,nm pump is red detuned by $\Delta$
  from the intermediate level $5\text{P}_{\sfrac{3}{2}},\,F=3$ to reduce the rate of incoherent scattering, with $\Delta$ between 30 and 60\,MHz.
  The two-photon detuning~$\delta$ is one of the parameters we study in this work.

The subsequent decay
from the excited level~\mbox{$5\text{D}_{\sfrac{3}{2}},F\!=\!3$}
to the ground state~\mbox{$5\text{S}_{\sfrac{1}{2}},F\!=\!2$}
via~\mbox{$5\text{P}_{\sfrac{1}{2}},F\!=\!2$}
generates a pair of photons
with wavelengths centered around 795\,nm (signal) and 762\,nm (idler).
We reject light originating from other scattering processes
using narrowband interference filters.
The geometry of the pump and collection modes is chosen to
satisfy the phase matching condition.
Energy conservation ensures time correlation of the generated photons, while
the time ordering imposed by the cascade decay results in a strongly asymmetrical time envelope of the biphoton.
This coherent process is accompanied by incoherent scattering.
Both processes generate light at the same wavelengths,
making it impossible to distinguish them by spectral filtering.
Similar to simple two-level systems \cite{Mollow_1969, cohen:2008},
coherent and incoherent scattering have different dependencies on a number of experimental parameters.

To understand the difference in behavior, we consider a long-established model
of a strongly driven three-level atom~\cite{Whitley:1976ej,Lawande:1986ch}.
This simple model correctly describes some of the features of our photon pair
source. In this model, the the atomic state is described by the $3\times 3$ density
matrix~$\rho$, where state~$1$ corresponds to the ground state, state 3 to the
most excited state, and state~$2$ to the intermediate state in the cascade decay.
The total scattering rate,
that includes both coherent and incoherent events,
is proportional to the population in state~3,
\begin{equation}\label{eq:chi33}
  r_{\mathrm{tot}} \propto \langle \rho_{33} \rangle\,,
\end{equation}
while the signal we are interested in is proportional to the coherence between
states 1 and 3,
\begin{equation}\label{eq:chi31}
  r_{\mathrm{coh}} \propto| \langle \rho_{31} \rangle |^2\,.
\end{equation}
Following~\cite{Whitley:1976ej}, we derive an analytical steady state solution
of the master equation as function of the pump intensities (through the
corresponding Rabi frequencies $\Omega_1$ and~$\Omega_2$) and detunings
($\Delta$ and ~$\delta$)~\footnote{These analytical forms are long and
  cumbersome, we have included them in the appendix.  Note that the solutions
  presented in~\cite{Whitley:1976ej} contain a mistake, as already pointed out
  by~\cite{Akimov:2010hm}}.

In order to compare Eq.~(\ref{eq:chi33}) and Eq.~(\ref{eq:chi31}) to our experimental results, we need to take into account the linewidths of the pump lasers.
A rigorous approach would require the inclusion of the laser linewidth in the master equation~\cite{McDonnell:2004db}.
For large Rabi frequencies, as in our case, the spectral broadening associated with the laser power dominates.
We can therefore approximate the combination of the two pump lasers
Lorentzian profiles of width~$\approx 1$\,MHz
into a single
noise spectrum with Gaussian profile~$G(\delta)$ of width~$\approx 2$\,MHz.
We obtain a fitting function for our results by convolving Eq.~(\ref{eq:chi33}) and~(\ref{eq:chi31}) with the combined linewidth of the pump lasers,
\begin{equation}\label{eq:single_th}
  r_{single}\propto r_{\mathrm{tot}}(\Omega_1, \Omega_2, \Delta, \delta) * G(\delta)\,,
\end{equation}
and
\begin{equation}\label{eq:pairs_th}
  r_{pairs}\propto r_{\mathrm{coh}}(\Omega_1, \Omega_2, \Delta, \delta) * G(\delta)\,.
\end{equation}
The heralding efficiency for photons (in a scenario where one photon
is used as a herald for the presence of the other) is the ratio of these
rates:
\begin{equation}\label{eq:eff_th}
  \eta = \frac{r_{pairs}}{r_{single}} = \frac{r_{\mathrm{coh}}(\Omega_1, \Omega_2, \Delta, \delta) * G(\delta)}{r_{\mathrm{tot}}(\Omega_1, \Omega_2, \Delta, \delta) * G(\delta)}\,.
\end{equation}

This model does not take into account the Zeeman manifold of the energy levels,
nor the collective interaction within the atomic ensemble.
We already presented a model and experimental evidence
of the effects of polarization choice for
pumps and collection modes previously~\cite{Gulati:2015ee}.
In the rest of this article,
the polarization of the pump beams and collection modes is chosen
to maximize the effective nonlinearity and, consequently, maximize the generation rates.
To understand the effect of collective interaction in a cascaded decay process we compare our results with the model proposed in~\cite{Jen:2012} in section~\ref{sec:OD}.

\section{Experimental setup}

  The experimental setup is shown in Fig.~\ref{fig:setup}(c).
  The non-linear medium is
  an ensemble of~$^{87}$Rb atoms in a vacuum chamber (pressure~$1\times10^{-9}$~mbar),
  trapped and cooled with a Magneto-Optical trap (MOT)
  formed by laser beams red detuned by
  24\,MHz
  from the cycling transition~$5\text{S}_{\sfrac{1}{2}},F\!=\!2\rightarrow5\text{P}_{\sfrac{3}{2}},F\!=\!3$,
  with a diameter of 15\,mm and an optical power of 45\,mW per beam.
  An additional laser tuned to the~$5\text{S}_{\sfrac{1}{2}},F\!=\!1\rightarrow5\text{P}_{\sfrac{3}{2}},F\!=\!2$
  transition optically pumps the atoms back into the~$5\text{S}_{\sfrac{1}{2}},F\!=\!2$
  level.
  The low temperature of the ensemble ensures a negligible Doppler broadening of the atomic transition line,
  resulting in a reduction of the bandwidth of the generated photons by an order of magnitude compared to the hot vapor sources~\cite{Willis:2010gp, Ding:2012gh}.

  In its initial implementation~\cite{Srivathsan:2013fa}, the source was non-collinear, i.e., pump and collection modes do not lie on the same axis.
  This approach was chosen to minimize the collection of any pump light into the parametric fluorescence modes.
  In subsequent experiments, including this work,  we instead chose a collinear configuration.
  This geometry simplifies the alignment
  and allows for a more efficient coupling
  of the generated photons into single mode fibers.
  We combine the pump beams (780\,nm and 776\,nm) using a narrowband interference filter (IF$_1$) as a dichroic mirror.
  Similarly, we separate the signal (762\,nm) and idler (795\,nm) modes using another interference filter (IF$_2$).
  Leaking of pump light into the collection modes is reduced by
  an additional
  interference filter
  in each collection mode (IF$_3$,  IF$_4$).
  All interference filters used in the setup have a full width half maximum
  bandwidth of 3\,nm and a peak transmission~$96\%$ at 780\,nm.
  We tune
  their transmission frequencies
  by adjusting the angles of incidence.
  Polarizers P$_1$ and P$_2$ fix the polarization of the fluorescence
  before collecting it
  into single mode fibers with aspheric lenses.
  Single photons are detected using avalanche photo diodes (APD) with quantum efficiency of $\approx50\%$.

  Fig.~\ref{fig:setup}(b) shows the timing sequence used in the experiment:
  16\,ms of cooling of the atomic vapors, followed by a 1\,ms time window,
  during which the cooling beams are off and pump~1 and pump~2 shine on the cloud.
  We use external-cavity laser diodes (ECDL) with bandwidths in the order of 1\,MHz to generate the pumps,
  and control their power and detuning using acousto-optic modulators (AOM).

\section{Detection of photon pairs}
    \begin{figure}
      \centering
      \includegraphics[width=\columnwidth]{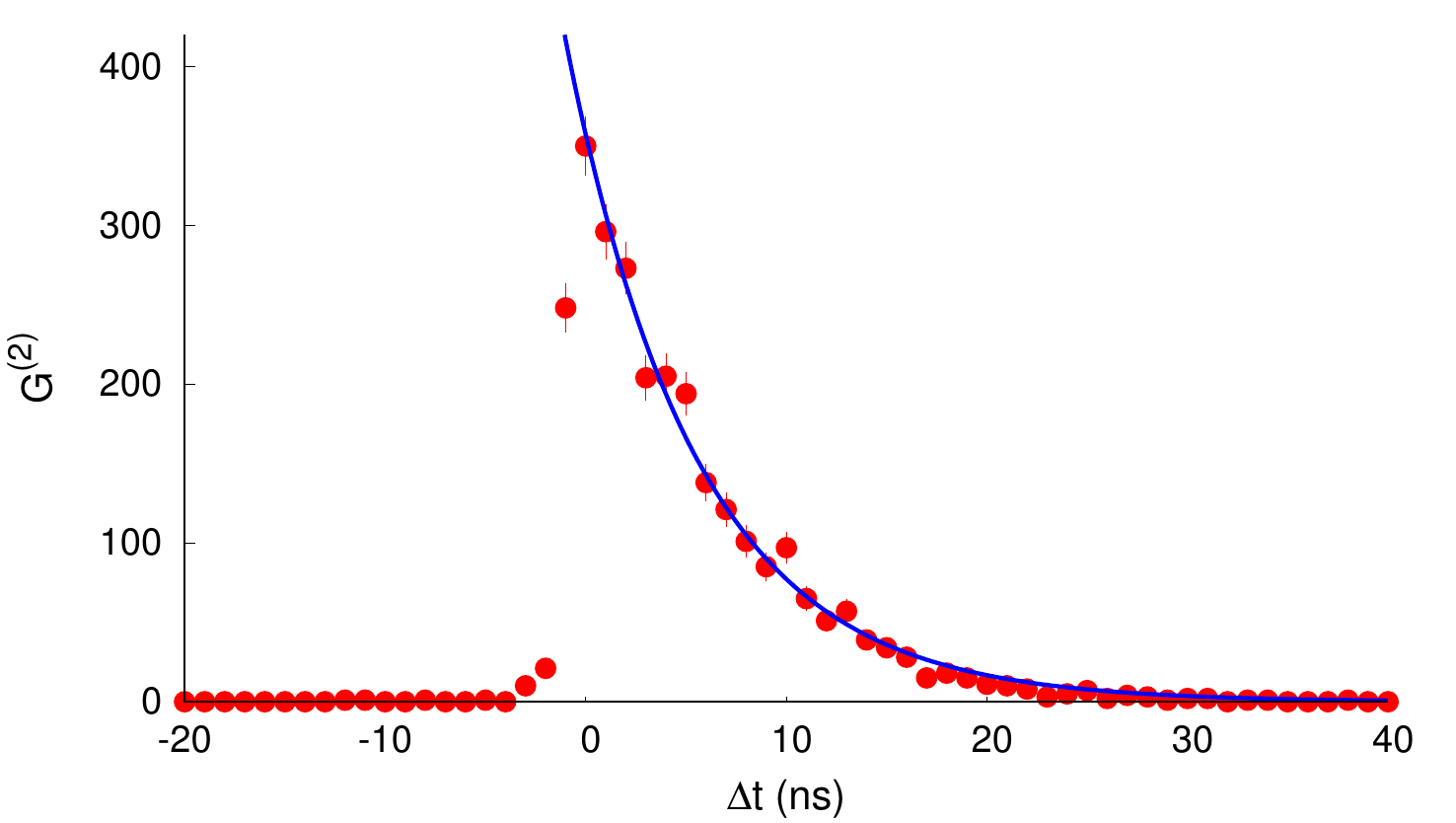}
      \caption{\label{fig:correlation_fn}
        Histogram of coincidence events~$G^{(2)}(\Delta t)$
        as a function of the time difference between the detection of signal and idler photons for a
        total integration time of~42\,s.
        Pump powers:~$P_{780}=450\,\mu$W and~$P_{776}=3$\,mW,
        detunings~$\Delta=-60 $\,MHz and $\delta=12$\,MHz.
        The solid line is a fit to the model described by Eq.~\ref{eq:g2_fit}, giving a value of~$\tau=6.52\pm0.04$\,ns.
        }
    \end{figure}

  We characterize the properties of the source from the statistics and correlation of detection times for events in the signal and idler modes.
  All detection events are timestamped with a resolution of 125\,ps.
  Fig.~\ref{fig:correlation_fn} shows
  a typical coincidence histogram~$G^{(2)}$,
  i.e., the coincidence counts as a function of the delay between detection times~$\Delta t$.
  The correlation function shows an asymmetric shape:
  a fast rise followed by a long exponential decay.
  The rise time is limited by the jitter time of the APDs (typical value~$\approx800$\,ps),
  while the decay is a function of the coherence time.
  In a previous work~\cite{Srivathsan:2013fa}
  we showed that
  the bandwidth is inversely proportional to the decay time constant~$\tau$.
  We measure~$\tau$ by fitting the histogram~$G^{(2)}$ with the function
  \begin{equation}\label{eq:g2_fit}
    G^{(2)}_{\text{fit}}(\Delta t)= G_{\text{acc}} + G_0\,e^{-\Delta t/\tau} \Theta (\Delta t)\,,
  \end{equation}
  where~$G_{\text{acc}}$ is the rate of accidental coincidences,~$\Theta$ is
  the Heaviside step function, and $G_0$ an amplitude.
  The rate of accidental coincidences $G_{\text{acc}}$ is fixed by considering
  the average of~$G^{(2)}$ for times $\Delta t$ much larger than the coherence time, leaving as free parameters only~$G_0$ and~$\tau$.

  To characterize the source, we consider the
  rate of single event detection in
  the signal ($r_s$) and idler ($r_i$) modes,
  together with the rate of coincidence detection
  ($r_p$) as
  the signature of photon pairs.
  All reported rates are instantaneous rates in the parametric conversion part
  of the cooling/photon generation cycle.

  The total pair detection rate $r_p$ of the source is obtained by
  integrating $G^{(2)}(\Delta t)$ over a coincidence time window
  $0 < \Delta t < \Delta t_c$.
  We choose
  $\Delta t_c=30$\,ns to ensure the collection of a large fraction of events also for the largest coherence times~$\tau$ observed.

  Another parameter we extract from the measured $G^{(2)}(\Delta t)$ is heralding efficiency.
  Due to the intrinsic asymmetry of the process we
  define a two heralding efficiencies from the same measurement,
  one for the signal,
  \begin{equation}\label{eq:eff_s}
    \eta_{\text{S}} = r_p/(r_{\text{S}}-d_{\text{S}})\,,
  \end{equation}
  and one for the idler,
  \begin{equation}\label{eq:eff_i}
    \eta_{\text{I}} = r_p/(r_{\text{I}}-d_{\text{I}})\,,
  \end{equation}
  where~$d_{\text{S}} = 508$\,s$^{-1}$ and~$d_{\text{I}} = 165$\,s$^{-1}$
  are the dark count rates on the signal and idler detectors.

\section{Effect of the number of atoms}\label{sec:OD}
  One of the parameters of interest is
  the number of atoms~$N$ participating in the four-wave mixing process.
  We control it
  by varying the optical power of the repump light during the cooling phase,
  thus changing the atomic density without altering the geometry of the optical trap.

  We estimate~$N$ by measuring
  the optical density~(OD) of the atomic ensemble
  for light resonant with the~$5\text{S}_{\sfrac{1}{2}},F\!=\!2\rightarrow5\text{P}_{\sfrac{3}{2}},F\!=\!3$ transition.
  To obtain a reliable measure of the OD,
  we turn off pump~2 and
  set pump~1 to~10\,$\mu$W, more than~40 times lower than the saturation intensity of the transition of interest.
  We record the transmission of pump~1 through the vacuum cell
  for a range of values of~$\Delta$
  wide enough to capture the entire absorption feature,
  and normalize it to the transmission observed without the atomic cloud.
  We fit the measurement results with the expected transmission spectrum
  \begin{equation}
    T(\Delta) = \exp \left(-\OD \frac{\gamma^2}{\Delta^2 + \gamma^2}\right)\,,
  \end{equation}
  with $\gamma= 6.067$\,MHz and~OD as the only free parameter.
  From the size of the probe beam $w_0\approx450$\,$\mu$m, we estimate~$N$.
  We observed a minimum of~$N\approx1.5\times10^{7}$, corresponding to an OD~$\approx 7$,
  and a
  maximum of~$N \approx 6.3\times10^{7}$, OD~$\approx 29$.
  We expect the effective number of atoms participating in the FWM process to decrease during the measurement due to the heating caused by the intense pumps.

  \begin{figure}[ht]
    \centering
      \includegraphics[width=\columnwidth]{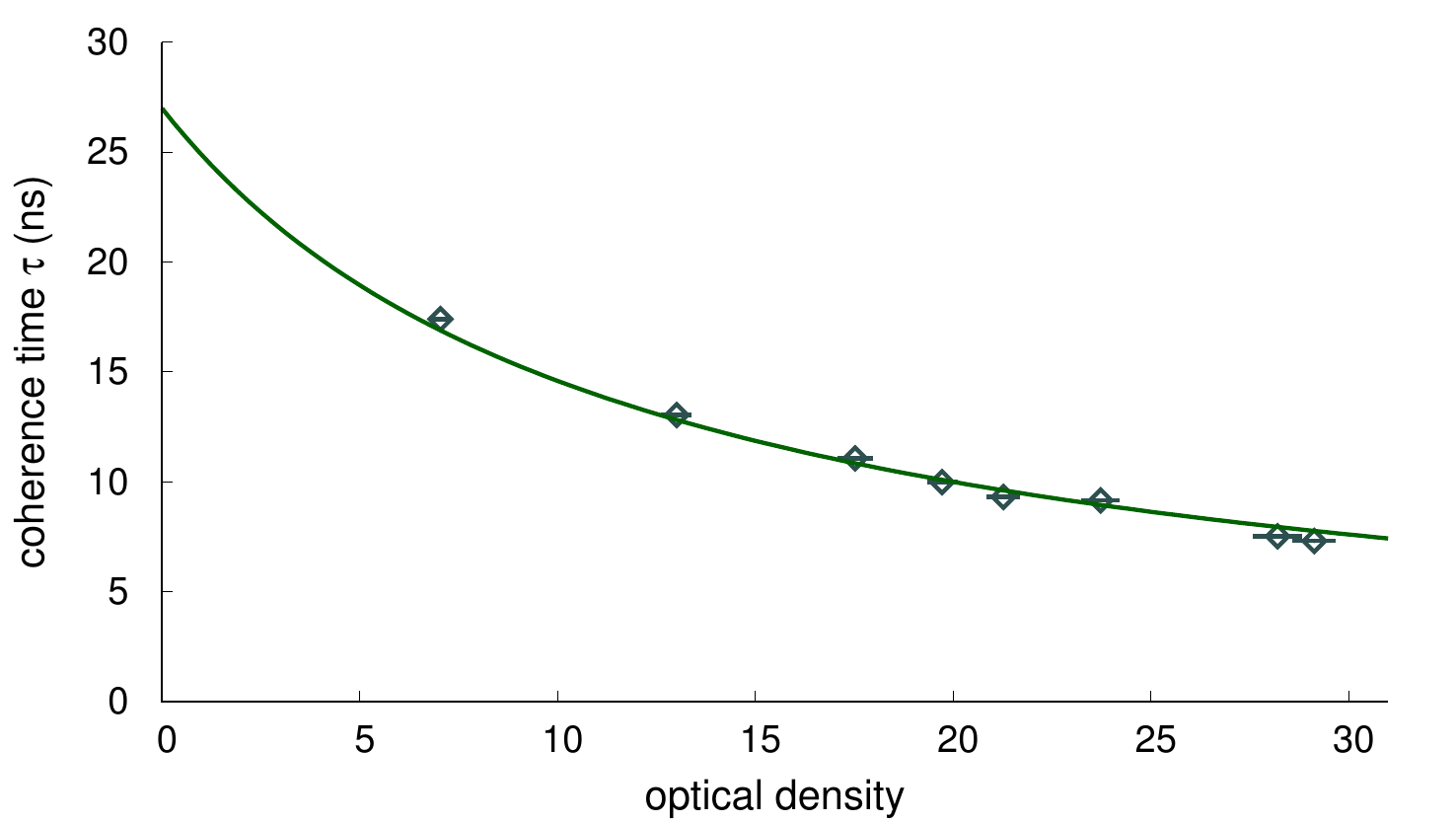}
      \caption{\label{fig:od_times}
      Coherence time of the photon pair
      as a function of the optical density (OD) of the atomic cloud.
      The solid line is obtained by fitting Eq.~\ref{eq:tau_od}, obtaining~$\mu=0.0827\pm0.002$.
      Other parameters:
      $P_{776} = 15$\,mW,
      $P_{780} = 300$\,$\mu$W,
      $\Delta=-60 $\,MHz,
      $\delta=12$\,MHz.
      }
  \end{figure}

\begin{figure}[ht]
    \centering
      \includegraphics[width=\columnwidth]{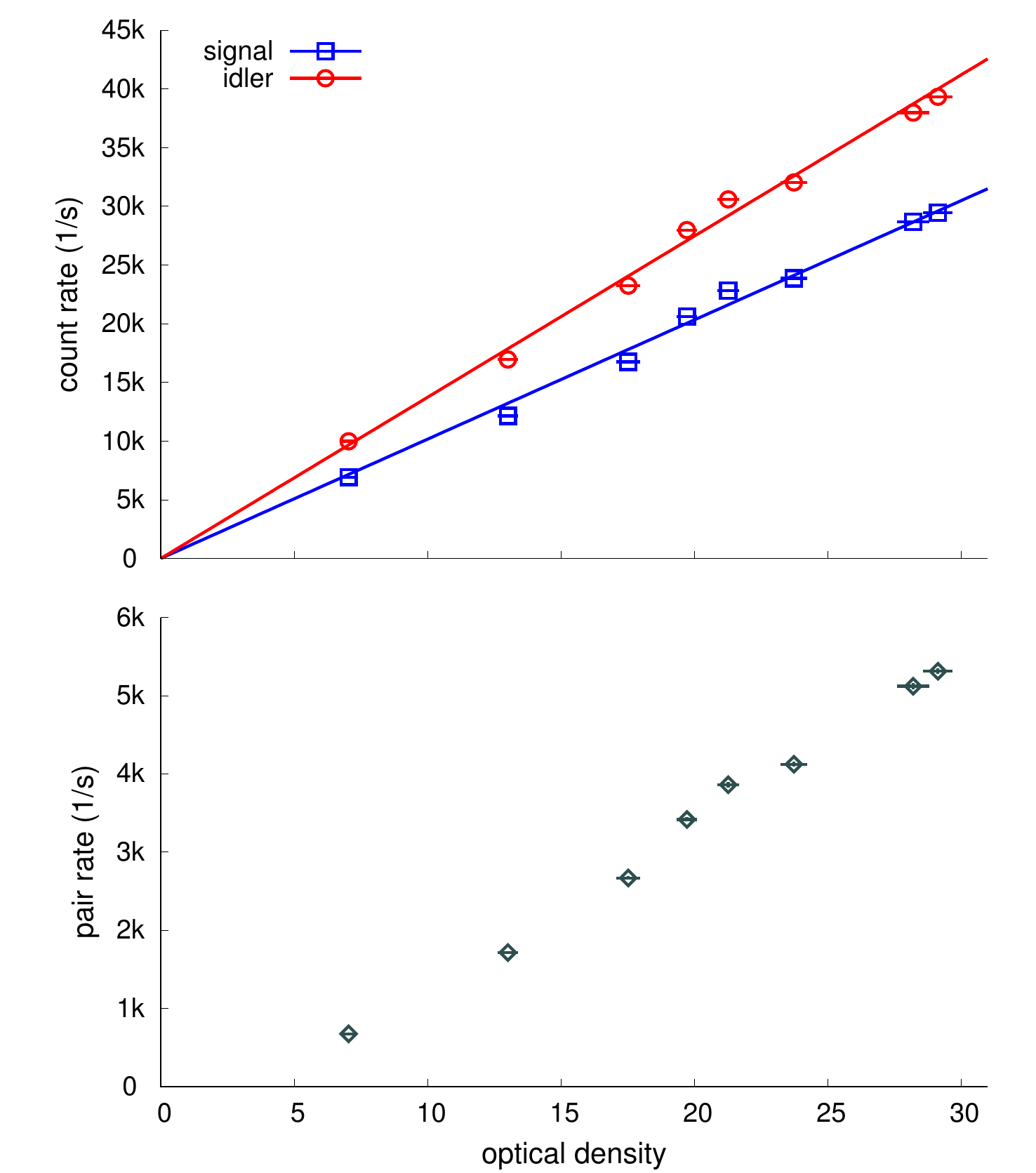}
      \caption{\label{fig:od_rates}
      Rate of single counts in the signal and idler modes (top),
      and rate of coincidence counts (bottom)
      as a function of the optical density (OD) of the atomic cloud.
      The solid lines are fits for $r_{s,i}= a_{s,i}\,\text{OD}$, with $a_{s,i}$ the only free parameter.
      Other parameters:
      $P_{776} = 15$\,mW,
      $P_{780} = 300$\,$\mu$W,
      $\Delta=-60 $\,MHz,
      $\delta=12$\,MHz.
      }
  \end{figure}

  Single detection rates for the signal~($r_s$) and idler~($r_i$) modes
  increase linearly with the number of atoms involved in the process,
  as expected for incoherent processes (see Fig.~\ref{fig:od_rates}).
  The increase of pair rate~$r_p$ with~$N$, however, appears to be faster than
  linear.

  Further, the decay or coherence time~$\tau$ decreases in our experiments as~OD
  increases (see Fig.~\ref{fig:od_times}).
  The measured coherence time is always shorter than the natural
  lifetime~$\tau_0$=27\,ns of the intermediate state
  expected for the spontaneous decay in free space of this transition to the
  ground state of $^{87}$Rb.
  This is a signature of collective effects in the cold atom cloud~\cite{Gross1982,Srivathsan:2013fa}.
  The solid line is a fit to the theoretical model proposed in~\cite{Jen:2012}:
  \begin{equation}\label{eq:tau_od}
    \tau=\frac{\tau_0}{1+\mu\,\text{OD}},
  \end{equation}
  where the free parameter~$\mu$ is a geometrical constant depending on the shape of the atomic ensemble.

  \begin{figure}[ht]
    \centering
      \includegraphics[width=\columnwidth]{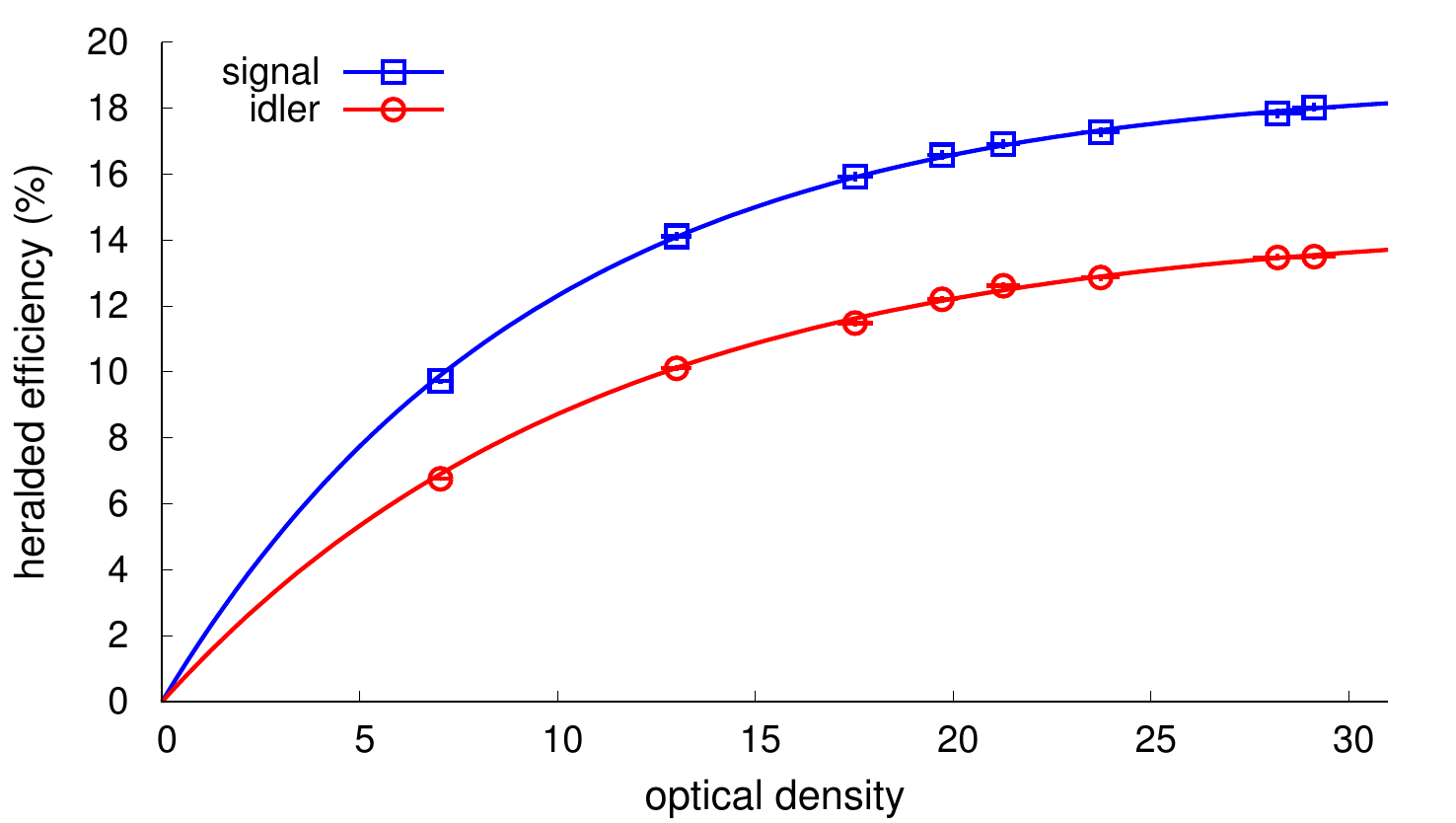}
      \caption{\label{fig:od_eff}
      Heralding efficiency for signal and idler modes as a function of the optical density.
      The solid lines are fits of Eq.~\ref{eq:eff_od}
      with~$\eta_{0s} = 0.190 \pm 0.001$ and~OD$_{0s} =9.7 \pm 0.1 $,
      and~$\eta_{0i} = 0.150 \pm 0.001$ and~OD$_{0i} =11.3 \pm 0.2 $.
      Other parameters:
      \mbox{$P_{776} = 15$\,mW},
      \mbox{$P_{780} = 300$\,$\mu$W},
      \mbox{$\Delta=-60$\,MHz},
      \mbox{$\delta=12$\,MHz}.
      }
  \end{figure}

  We do not have a complete explanation for the nonlinear increase of the pair
  rate with the optical density, but some insight can be gained from the
  heralding efficiencies shown in Fig.~\ref{fig:od_eff}.
  Both heralding efficiencies~$\eta_{s}$ and~$\eta_{i}$
  exhibit a saturation behavior that is described by the relation
  \begin{equation}\label{eq:eff_od}
    \eta_j=\eta_{0j}\left[1-\exp\left(-\frac{\text{OD}}{\text{OD}_{0j}}\right)\right]\quad\text{with}\quad j=s,i\,,
  \end{equation}
  where~$\eta_{0j}$ and~OD$_{0j}$ are free parameters.
  This heuristic expression suggests that (a) a higher optical density of
  the atomic cloud leads to an increase of the pair rate at the expense of a
  larger photon bandwidth, and (b) for large enough OD there is no improvement of
  heralding efficiency. Such considerations are discussed in
  section~\ref{sec:guideline}.

\section{Rates and heralding efficiencies}
  \begin{figure}
      \centering
      \includegraphics[width=\columnwidth]{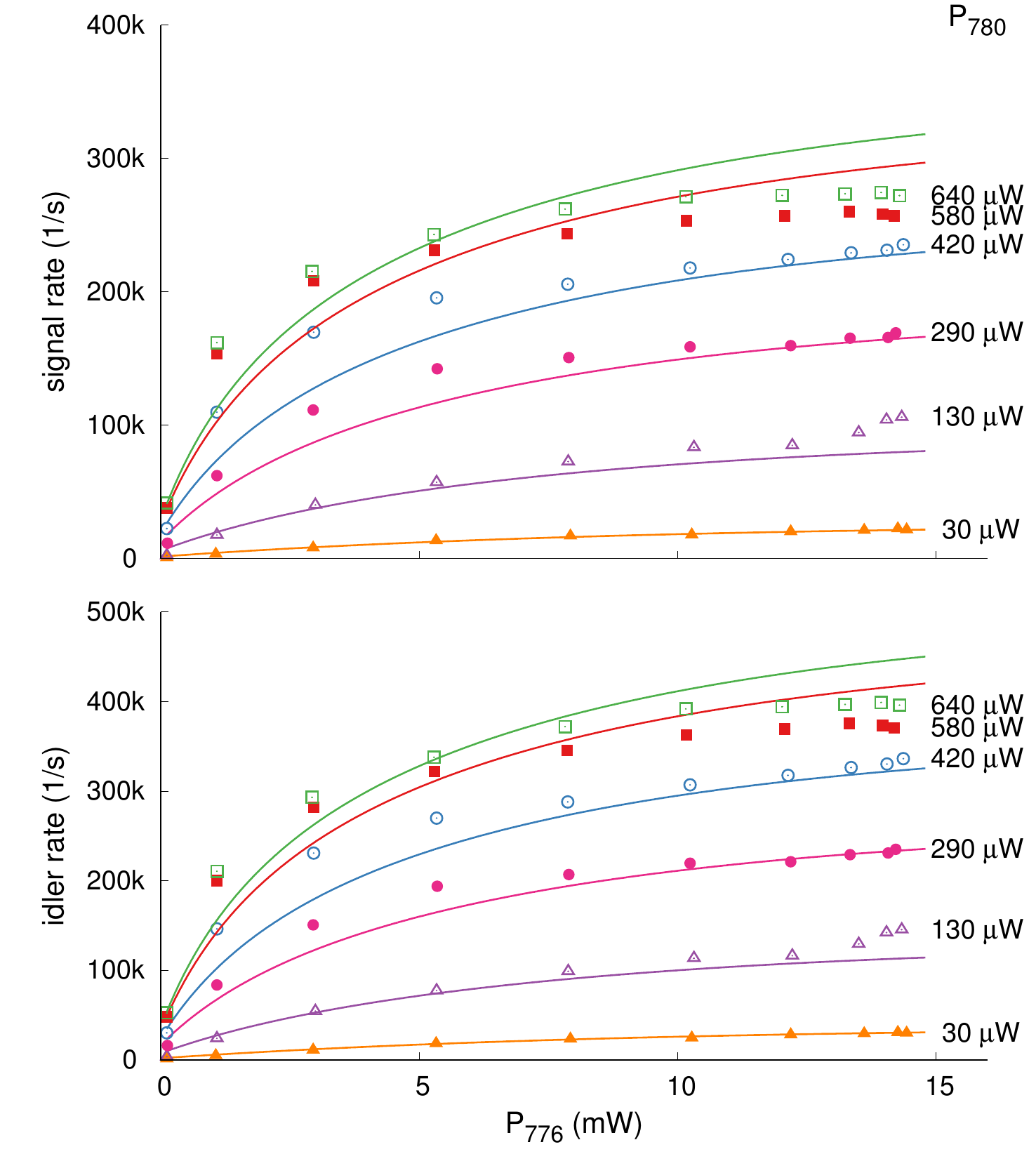}
      \caption{\label{fig:power_rates}
      Single rates for the signal (top) and idler (bottom) as a function of  pump power at~776\,nm ($P_{776}$) for different pump powers at~780\,nm.
      The vertical error bar on each point is smaller than the size of the data points.
      Other parameters:
      $\text{OD}=29$,
      $\Delta=-60 $\,MHz,
      $\delta=3$\,MHz.
      The solid lines are numerical fits with Eq.~\ref{eq:single_th}.
      }
  \end{figure}

\begin{figure}[ht]
      \centering
      \includegraphics[width=\columnwidth]{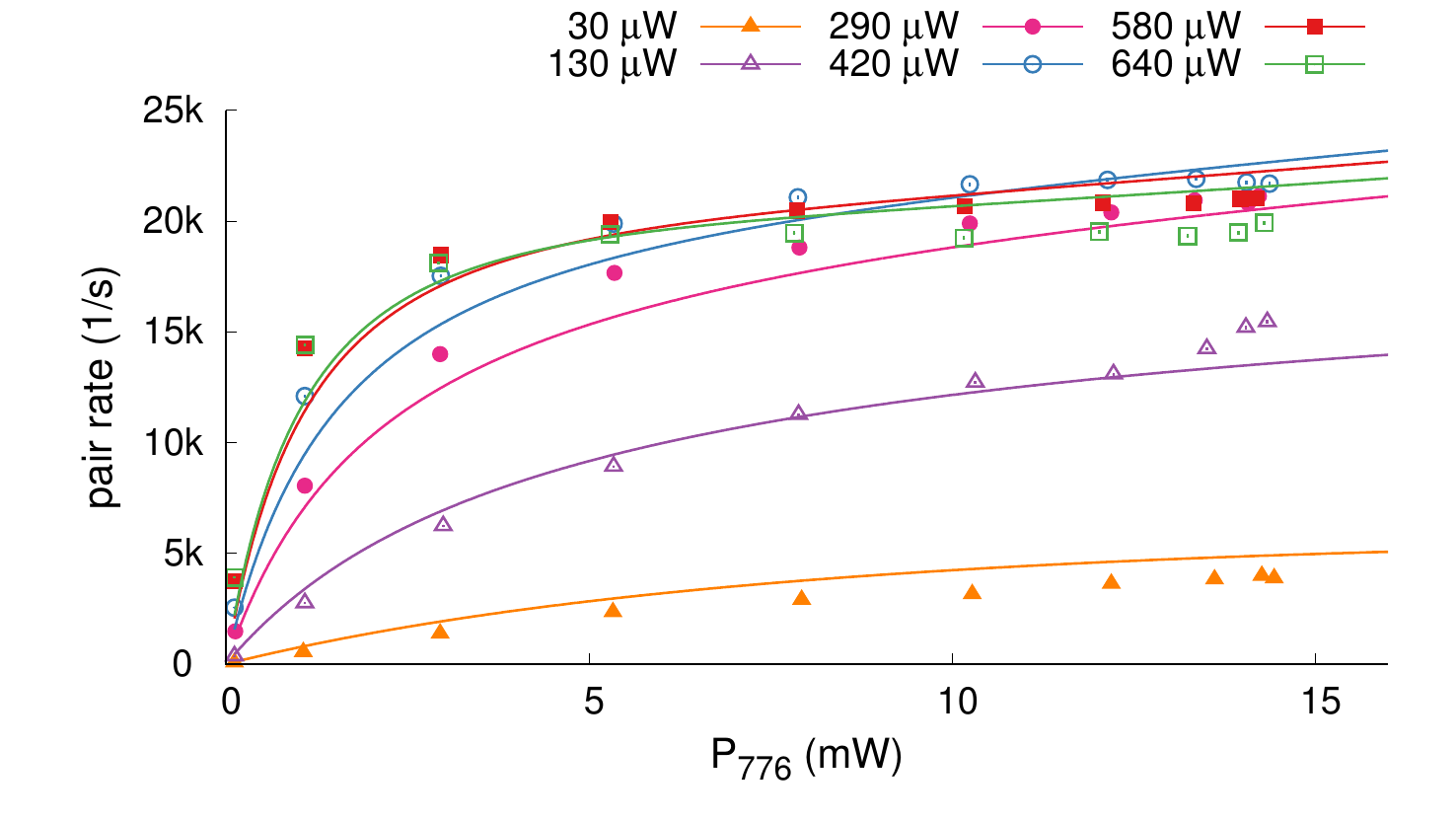}
      \caption{\label{fig:power_pairs}
      Pair rates
      as function of  pump power at 776\,nm ($P_{776}$) for different pump powers at 780\,nm.
      The vertical error bar on each point is smaller than the size of the data points.
      The solid lines are calculated from the theory.
      Other parameters:
      $\text{OD}=29$,
      $\Delta=-60 $\,MHz,
      $\delta=3$\,MHz.
      The solid lines are numerical fits with Eq.~\ref{eq:pairs_th}.
      }
  \end{figure}

  Brightness, a common parameter to characterize a photon pair source,
  is defined as
  the experimentally accessible rate of photon pairs emitted into the desired
  modes per mW of pump power.
  In our source, saturation effects of the atomic transitions involved give
  rise to a non-linear correlation between pump power and rates.
  In Fig.~\ref{fig:power_rates} and~\ref{fig:power_pairs}, the
  instantaneous single rates, $r_s$ and~$r_i$, and pair rates~$r_p$
  as a function of power in both pump transitions are shown.

  For a fixed two-photon detuning~$\delta$,
  all rates exhibit a saturation behavior.
  This suggests that an increase of the pump powers will increase the observed
  pair rate only to some extent, and an increased number of atoms of the
  ensemble might be a better option. However, as discussed in the previous
  section, this comes at the expense of a larger bandwidth.
  We also note that, while the model introduced in section~\ref{sec:intro}
  qualitatively explains the saturation behavior with the pump powers, it
  does not capture well the experimental observation for high powers. This is
  probably due to the optical pumping caused by the intense pump beams, which
  is not part of the relatively simple model.

\begin{figure}[ht]
      \centering
      \includegraphics[width=\columnwidth]{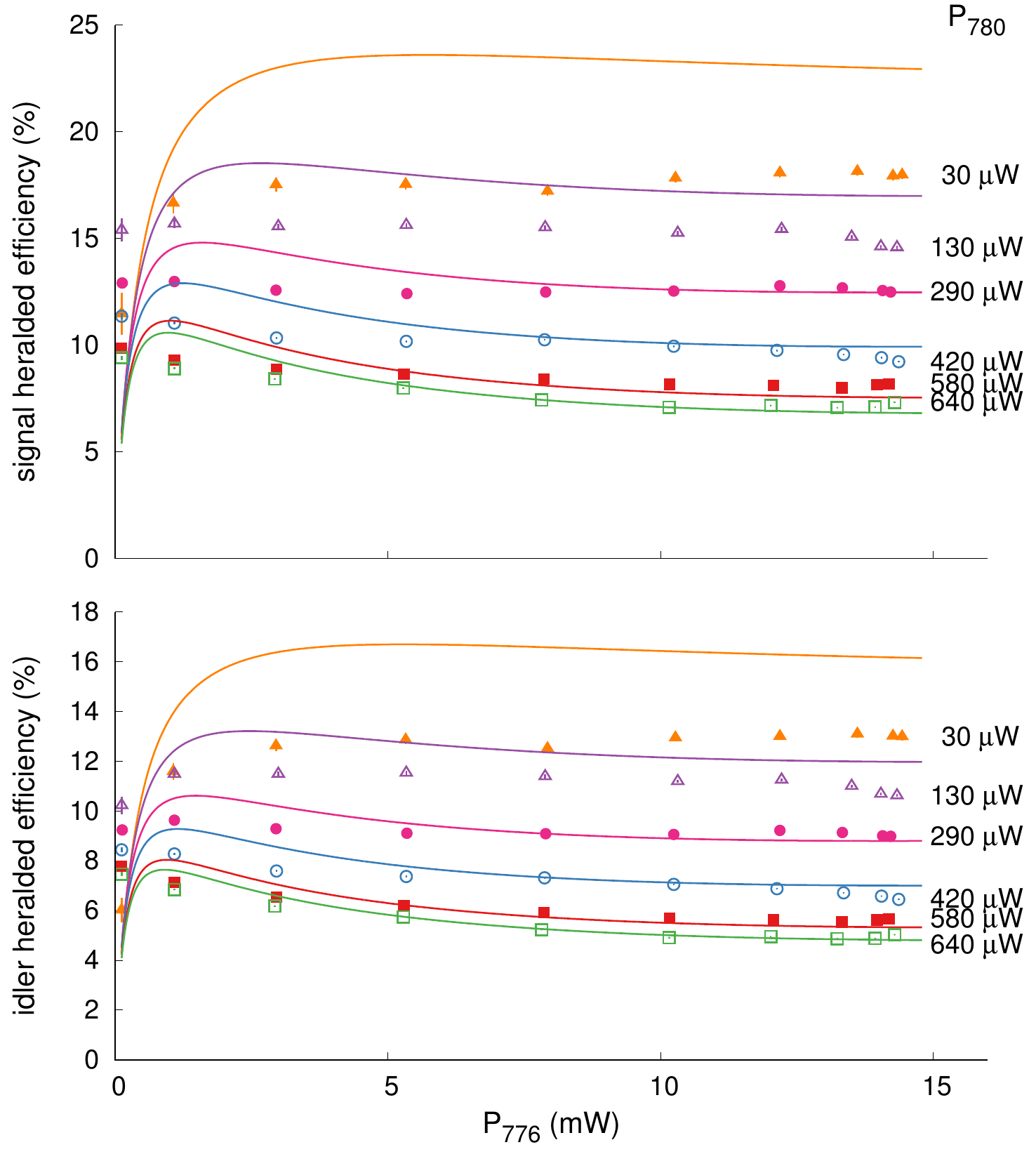}
      \caption{\label{fig:power_eff}
      heralding efficiency as function of~$P_{776}$ for the signal (top) and idler (bottom) for different~$P_{780}$.
      The vertical error bar on each point is smaller than the size of the data points.
      Other parameters:
      OD=29,
      $\Delta=-60 $\,MHz,
      $\delta=3$\,MHz.
      The solid lines are a numerical fit with Eq.~\ref{eq:eff_th}.
      }
  \end{figure}

  The dependency of heralding efficiencies on both pump powers is shown in
  Fig.~\ref{fig:power_eff}, both for our experimental observations and the
  model predictions.

  The intuition of a higher heralding efficiency at low pump powers due to a
  smaller contribution from incoherent processes is both found in the
  experiment and predicted by the model, but the model does not match the
  observations at low powers very well.
  A possible explanation is in one of the assumptions of our model.
  For low pump powers, the broadening due to Rabi frequencies of the pumps is comparable with the pump lasers linewidths, requiring then a different approach than convolution with a combined noise spectrum.
  However, our simple model ignores all geometrical aspects in
  the process, and therefore does not capture any spatial variation of the
  atomic density profile of the cloud, the intensity profile of the pump
  beams, or their respective overlap.

  Despite the limitations of the model, the observed power dependency of pair
  rates and heralding efficiency shown in Fig.~\ref{fig:power_pairs}
  and~\ref{fig:power_eff} suggest a strategy for optimizing the source
  brightness:
  a low power $P_{780}$ on the transition depopulating the ground state should
  ensure a high heralding efficiency, while a high power $P_{776}$ on the
  transition populating the state 3 should increase the brightness.
  An obvious experimental limitation to this strategy for Rubidium is the
  available~$P_{776}$.

  \begin{figure}[ht]
    \centering
    \includegraphics[width=\columnwidth]{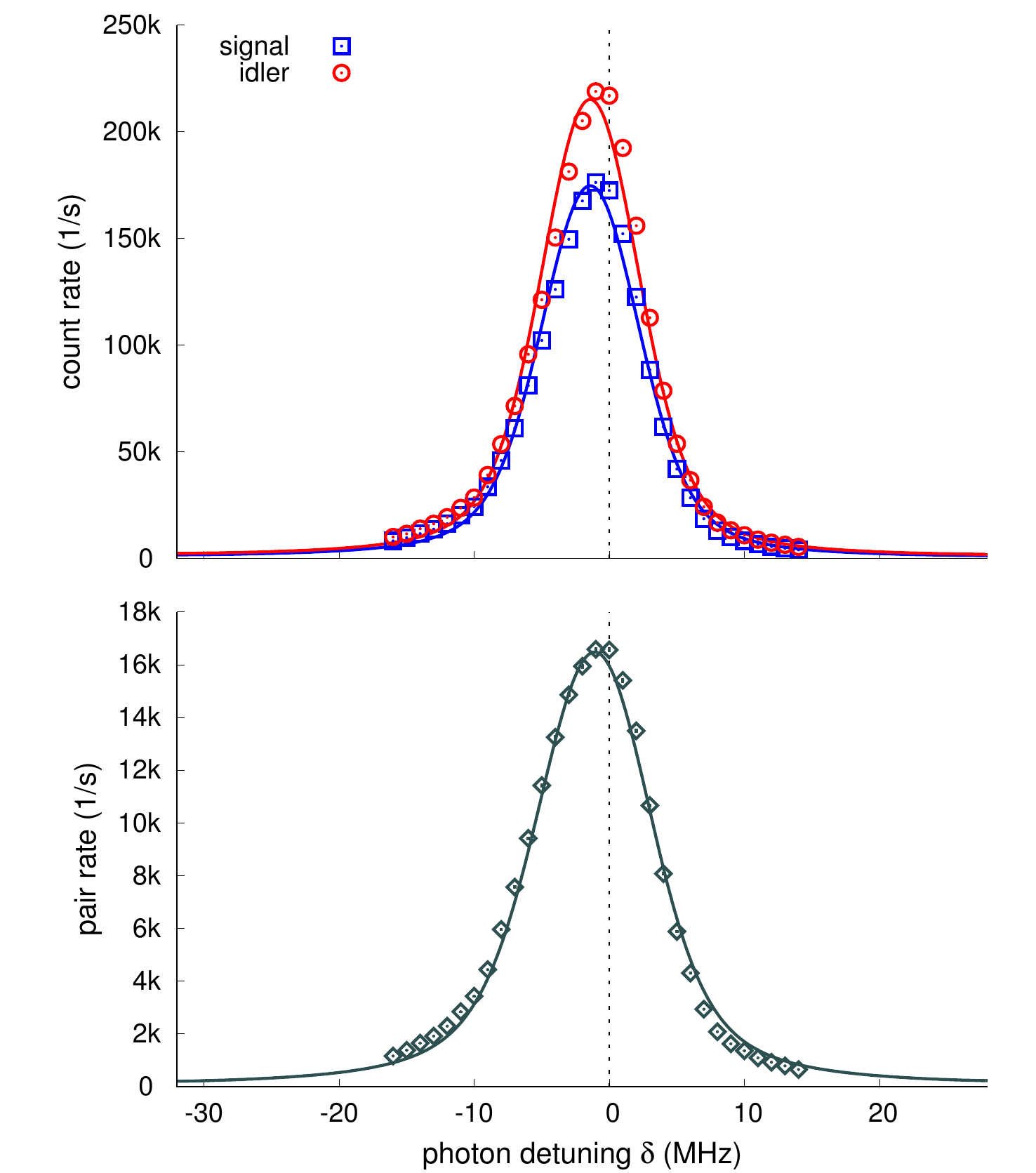}
    \caption{\label{fig:detuning_rates}
    (Top) Single count rates as a  function of the detuning from the two photon resonance~$\delta$.
    The solid lines are numerical fits of Eq.~\ref{eq:single_th}.
    (Bottom) Pair rate ($r_p$) as a function of~$\delta$.
    The solid line is a numerical fit of Eq.~\ref{eq:pairs_th}.
    Other parameters:
    $P_{776} = 15$\,mW,
    $P_{780} = 450$\,$\mu$W,
    \mbox{$\Delta=-60 $\,MHz},
    OD=29.
    The dotted line indicates~$\delta=0$.
    }
  \end{figure}

Apart from the optical power in the pump beams, other easily available
experimental parameters in the four wave mixing process are the pump
detunings.  Both single and pair rates have a strong dependence on the two-photon
detuning~$\delta$ from the ground state in the upper excited state, and have a
maximum at $\delta\approx0$, as expected for a scattering process (see
Fig.~\ref{fig:detuning_rates}).
The two-step nature of the excitation process leads to asymmetries in the
peaks, which is also predicted by the simple model of Eq.~(\ref{eq:single_th}) and~(\ref{eq:pairs_th}). To allow for a fair
comparison between the model prediction and the experimental data, we have
to take into account the linewidth of the pump lasers ($\approx 1 $\,MHz
each). We therefore convolve the theoretical predictions in Equations
\ref{eq:single_th} and \ref{eq:pairs_th} with a Gaussian
distribution modeling our laser noise. The resulting spectral profiles in the
two-photon detuning of pair and single rates then match very well the behavior
observed in our experiment.

\begin{figure}[ht]
    \centering
    \includegraphics[width=\columnwidth]{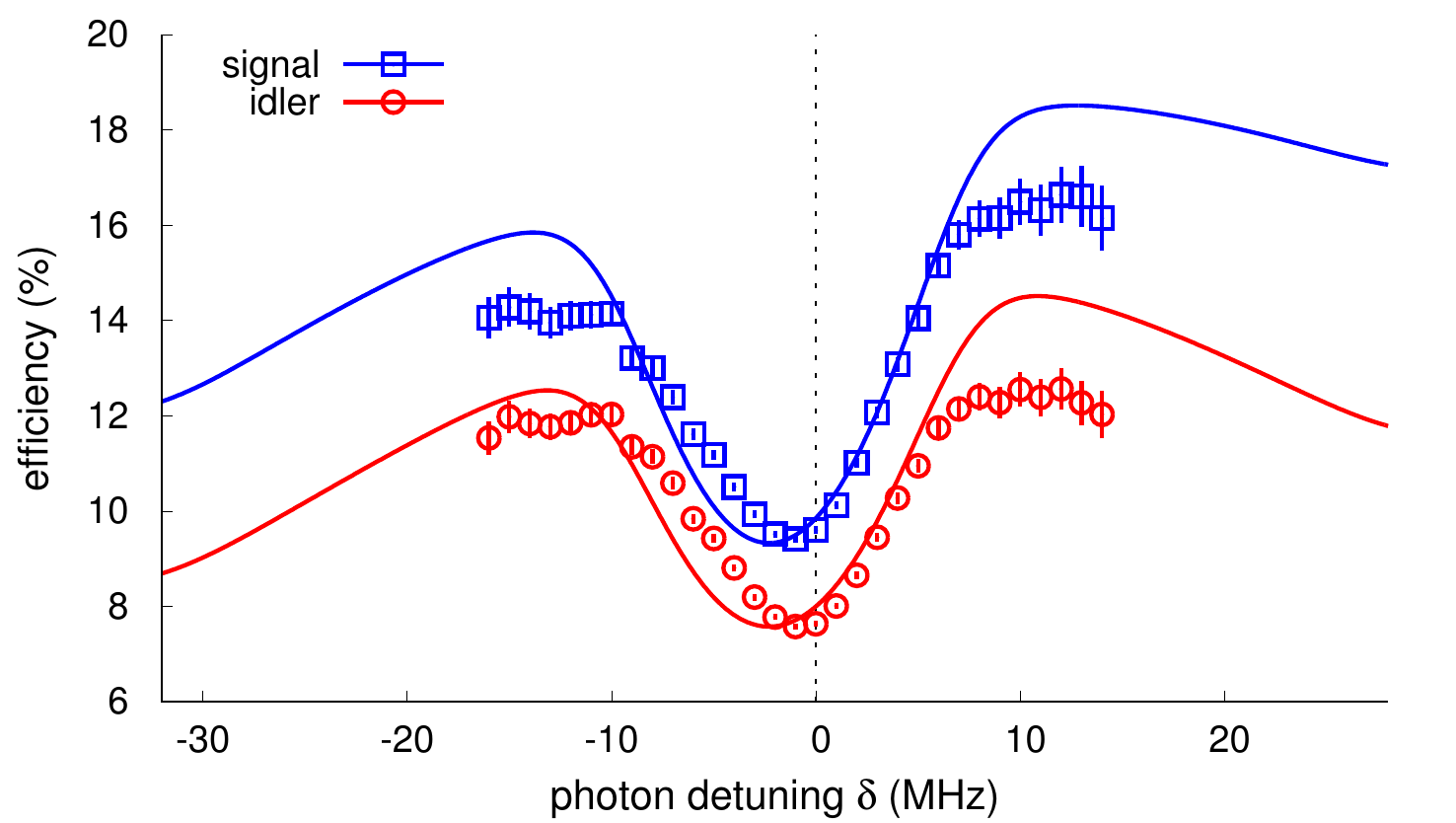}
    \caption{\label{fig:detuning_eff}
    Efficiency of the source as a function of the detuning from the two photon resonance~$\delta$.
    Other parameters:
    $P_{776}=15$\,mW,
    $P_{780}=450$\,$\mu$W,
    \mbox{$\Delta=-60$\,MHz},
    OD=29.
    The solid lines are fits with Eq.~\ref{eq:eff_th},
    the dotted line indicates~$\delta=0$.
    }
  \end{figure}

Contrary to the single and pair rates, both heralding efficiencies show an
asymmetric dip around $\delta\approx0$ (see Fig.~\ref{fig:detuning_eff}) in
our experiment, which is well captured by the model via Eq.~\ref{eq:eff_th}.

  This dip can be understood by taking into account that the observed single rate
  is the combination of FWM, a coherent process, and incoherent scattering,
  with the latter growing faster as~$\delta$ approaches~$0$.
  When choosing the operation parameter of a photon pair source for subsequent
  use, the two-photon detuning can therefore be optimized for a compromise
  between pair rate and heralding efficiency.

\section{Coincidence to accidental ratio (CAR)}
  \begin{figure}
    \centering
    \includegraphics[width=\columnwidth]{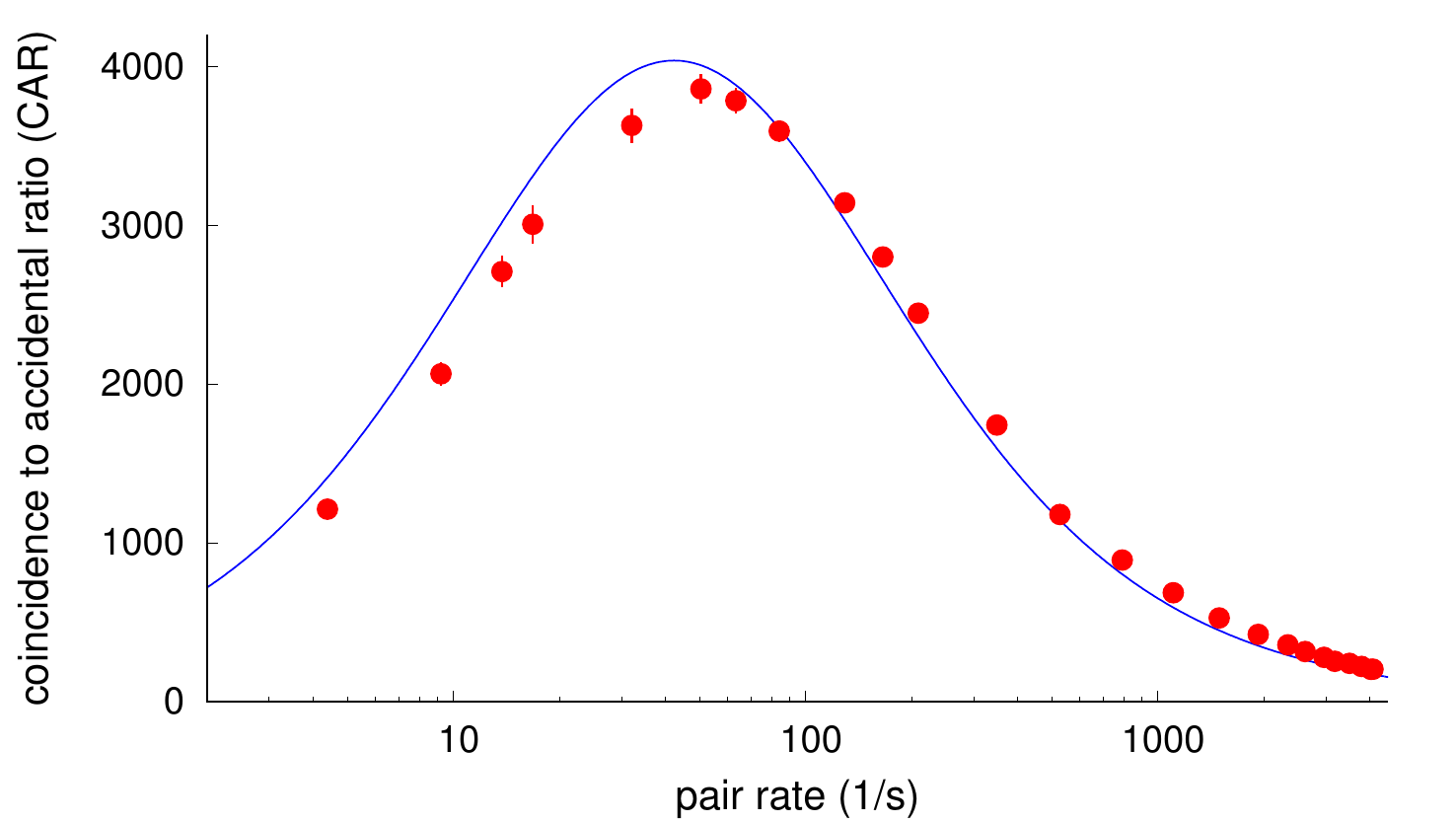}
      \caption{\label{fig:CAR}
      The coincidence to accidental ratio (CAR) as a function of pair rates~$r_p$.
      The solid line is obtained from Eq.~(\ref{eq:CAR_model}) with~$\eta_{\text{S}}=17.3\%$,~$\eta_{\text{I}}=12.4\%$,~$d_{\text{S}}=165$\,s$^{-1}$, $d_{\text{I}}=508$\,s$^{-1}$, $\Delta t=30$\,ns.
      }
  \end{figure}
Another relevant parameter for characterizing the usefulness of a source of photon pairs is
  the coincidence to accidental ratio~(CAR)~\cite{Takesue:2010gd,Xiong:2011kq},
  \begin{equation} \label{eq:CAR}
    \text{CAR}=\frac{R_p}{r_a}=\frac{r_{\text{I}}
        \, r_{\text{S}}  \, \Delta t + r_p}{r_{\text{I}}
        \, r_{\text{S}}  \, \Delta t }\,,
    \end{equation}
  where the accidental rate $r_a$ captures
  noise photons that degrade the correlation characteristics of
  the photon pair source.
  The connection between the CAR and pair rate~$r_p$ is shown in
  Fig.~\ref{fig:CAR}. In this parametric plot, we vary the pump power~$P_{776}$.
  Over a wide range of pair rates, the CAR increases when~$P_{776}$ is reduced because $r_a\appropto r^2_p$.
  For the experimental parameters shown in this measurement, the CAR peaks at~$\approx3800$, at a relatively low pair rate of~$r_p=50$\,s$^{-1}$.
  With a further reduction in pump power (and therefore in $r_p$), the CAR
  drops to~1, as background noise and detector's dark counts ($r_a$)
  dominate in Eq.~\ref{eq:CAR}.

To model the experimentally observed CAR, we modify the expression in
Eq.~\ref{eq:CAR} by separating the single rates for signal and idler into a
contribution from pairs, corrected by the respective heralding efficiencies,
and dark/background contributions for signal and idler. Signal and idler
heralding efficiencies vary very little over a wide range of pump
powers $P_{776}$, so we fix them to a single value. The resulting expression
for the CAR,
  \begin{equation} \label{eq:CAR_model}
    \text{CAR} =
    \frac{
      \left(\frac{r_p}{\eta_{\text{S}}} + d_{\text{S}} \right) \,
      \left(\frac{r_p}{\eta_{\text{I}}} + d_{\text{I}} \right) \,
      \Delta t + r_p}
      {
      \left(\frac{r_p} {\eta_{\text{S}}} + d_{\text{S}} \right) \,
      \left(\frac{r_p} {\eta_{\text{I}}} + d_{\text{I}} \right) \,
      \Delta t}\,,
  \end{equation}
reproduces very well the observed behavior in the experiment, suggesting
that the relation between CAR and pair rates is fairly well understood.

\section{Coherence time of the generated pairs}
\begin{figure}[ht]
    \centering
      \includegraphics[width=\columnwidth]{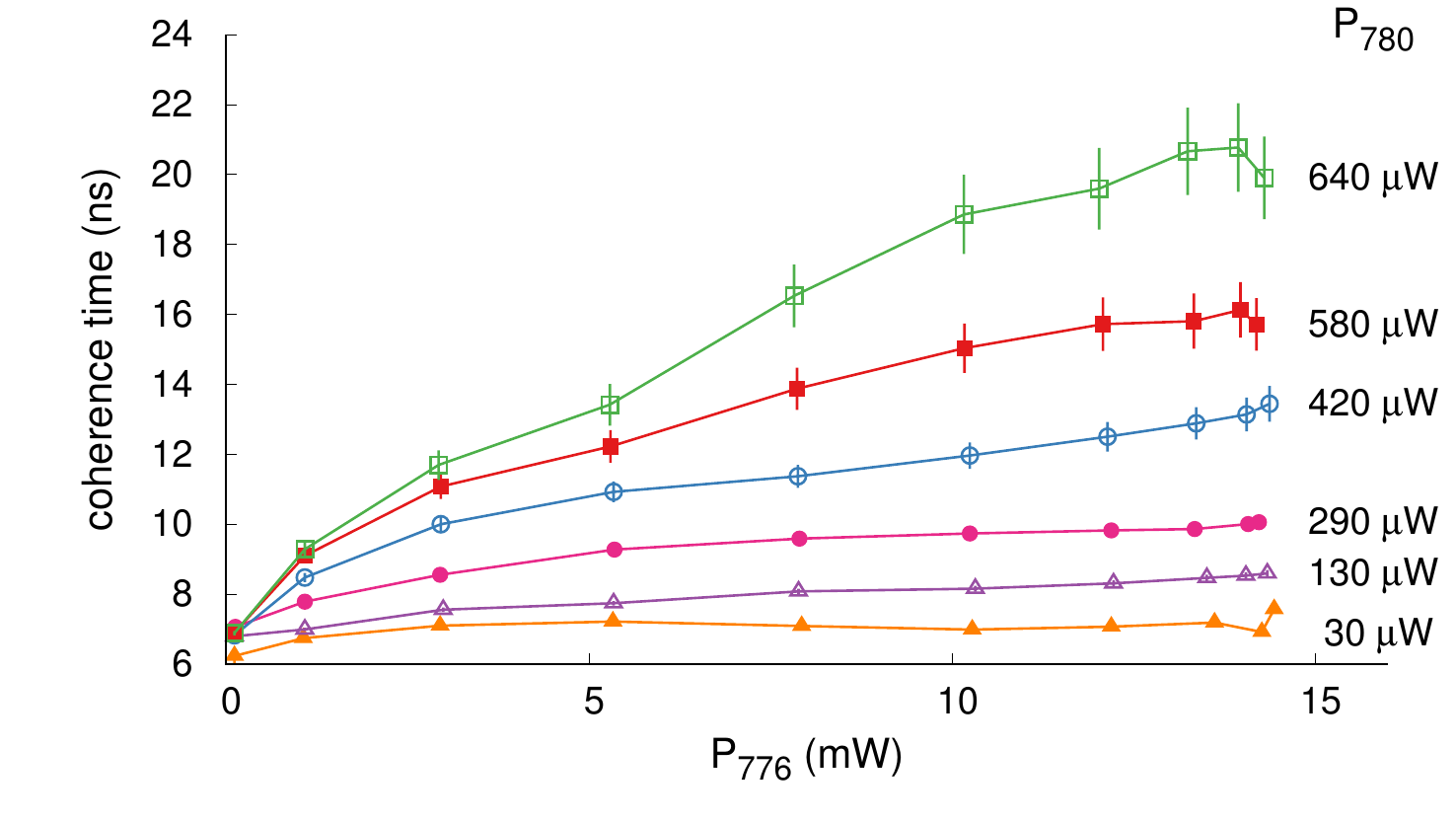}
      \caption{\label{fig:power_times}
      Coherence time as function of pump powers.
      Other parameters:
      $\text{OD}=29$,
      $\Delta=-60 $\,MHz,
      $\delta=3$\,MHz.
      }
  \end{figure}

\begin{figure}[ht]
    \centering
      \includegraphics[width=\columnwidth]{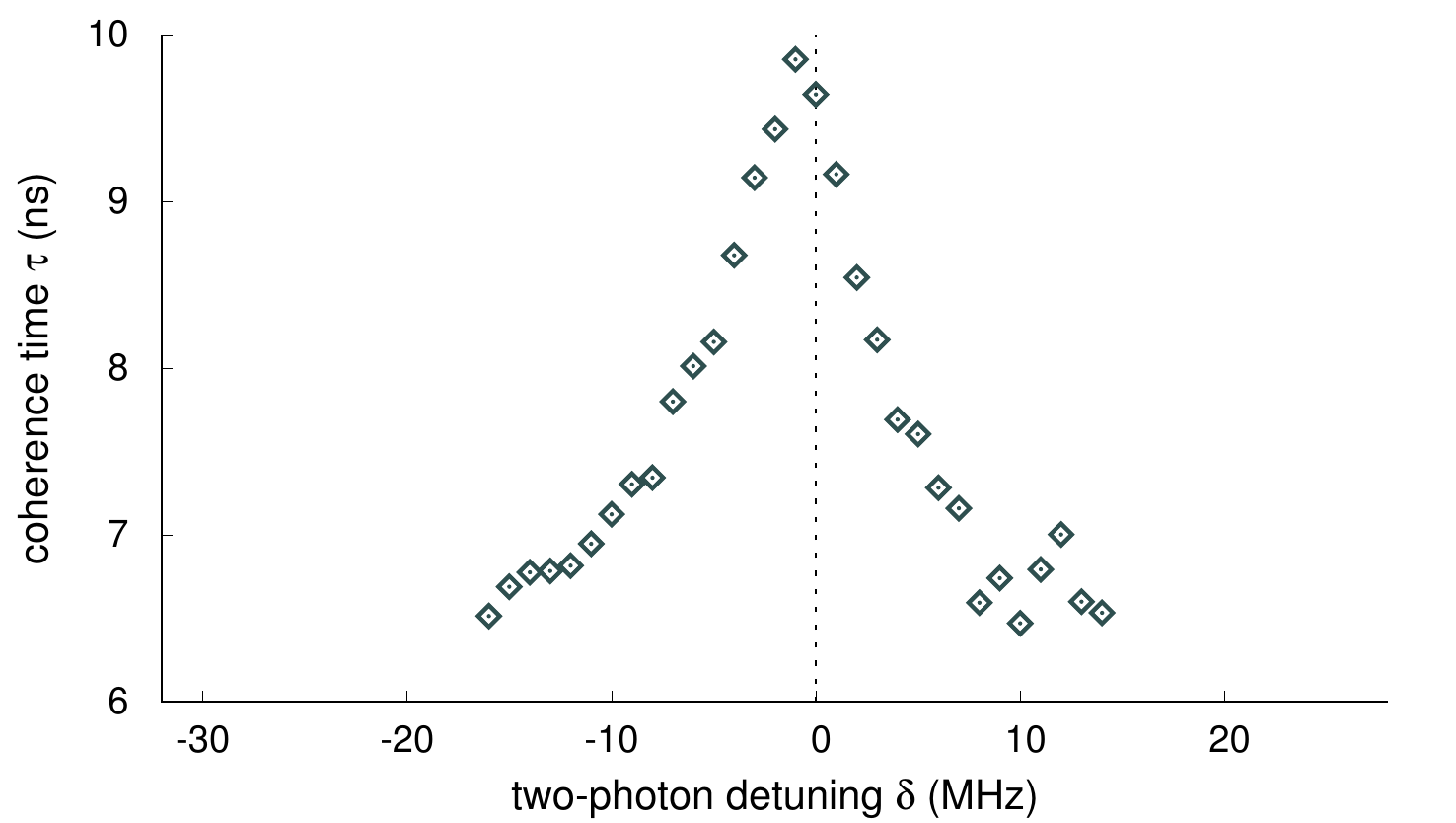}
      \caption{\label{fig:detuning_times}
      Coherence time as function of detuning.
      Other parameters:
      $P_{776}=15$\,mW,
      $P_{780}=450$\,$\mu$W,
      \mbox{$\Delta=-60$\,MHz},
      $\text{OD}=29$.
      The dotted line indicates~$\delta=0$.
      }
  \end{figure}

An important property of photon pair sources based on nonlinearities is the
small bandwidth of the emerging photons corresponding to a long
coherence time. The dependency of the coherence time, measured by fitting
photon pair timing histograms to Eq.~\ref{eq:g2_fit}, on pump power and
two-photon detuning is shown in Fig.~\ref{fig:power_times} and
\ref{fig:detuning_times}. The coherence time increases with both pump powers,
and also shows a maximum with respect to the two-photon detuning slightly below the
two-photon resonance, similar to the pair rates.

The simple 3-level model in section~\ref{sec:intro} does not address the
coherence time of the emerging photons. Even a more complex model
that includes the collective effects associated with the number of atoms~\cite{Jen:2012}
predicts only a dependency of
the coherence time on the number of atoms involved in the four-wave mixing
process (superradiance), but not on the pump power and two-photon detuning.
A possible reason for the observed dependency is a decay from
the excited state~$5\text{P}_{\sfrac{1}{2}},F\!=\!3$
to~$5\text{S}_{\sfrac{1}{2}},F\!=\!1$,
a ground state that does not participate in the coherent four wave mixing we are interested in,
effectively depleting the number of atoms interacting with the pump beams.
This depletion increases with pump intensities, and decreases with detuning,
and is not completely neutralized by the repump beam, resulting in a change of
the number of atoms in the participating ground state, which would then affect
the coherence time according to the more complex conversion
model~\cite{Jen:2012}.

To arrive at long coherence times, one therefore would need to optimize the
repumping process during the parametric conversion cycle in our experiment to
maintain the atomic population in the ground state.

\section{Guidelines for choice of parameters}\label{sec:guideline}
  \begin{figure}
    \centering
    \includegraphics[width=\columnwidth]{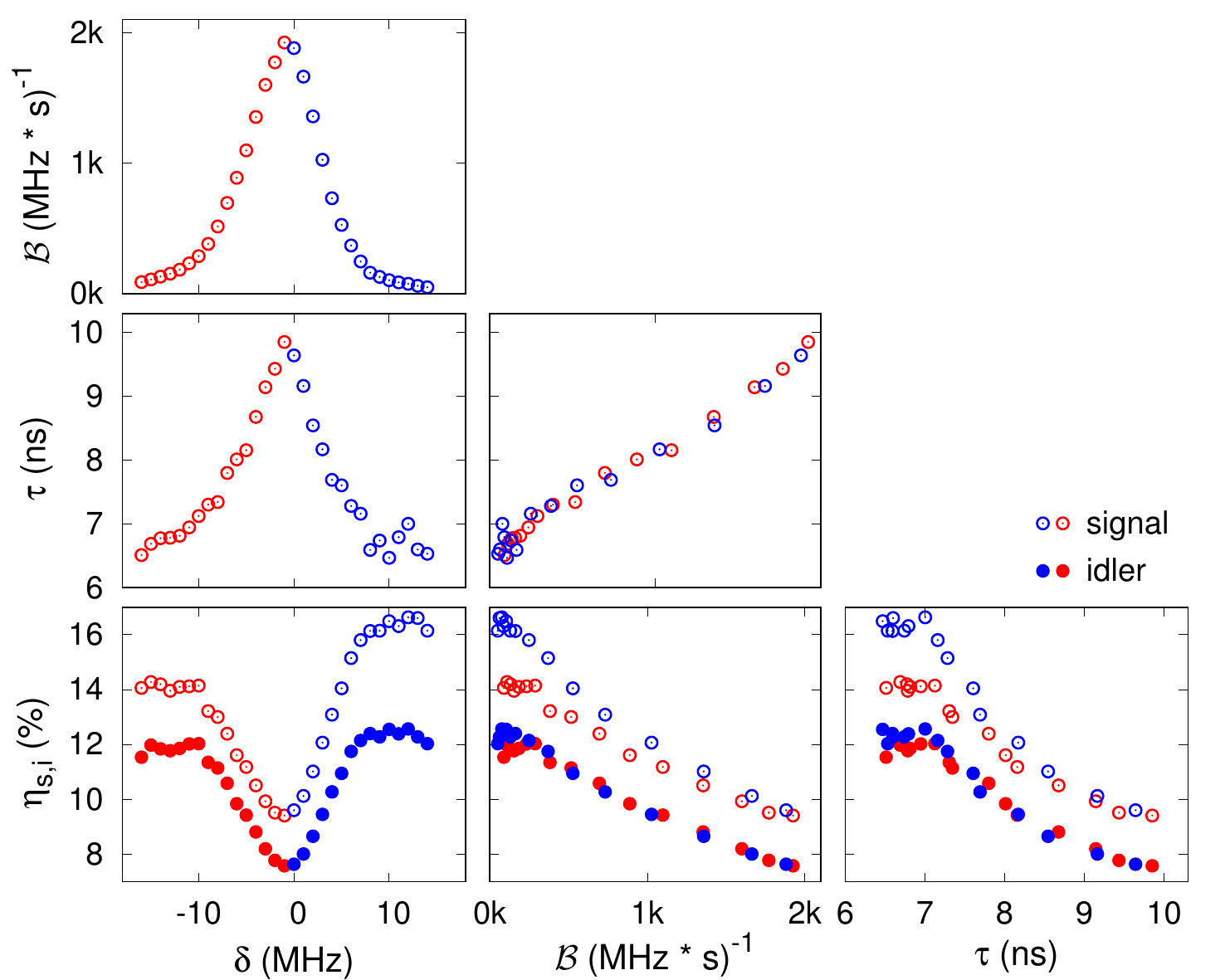}
    \caption{\label{fig:detuning_table}
    Summary of the effect of two-photon detuning~$\delta$ on heralding efficiencies~$\eta_{s, i}$, coherence time~$\tau$, and spectral brightness~$\mathcal{B}$.
    Other parameters:
    $P_{776}=15$\,mW,
    $P_{780}=450$\,$\mu$W,
    \mbox{$\Delta=-60$\,MHz},
    OD=29.
    }
  \end{figure}

  \begin{figure}
      \centering
      \includegraphics[width=\columnwidth]{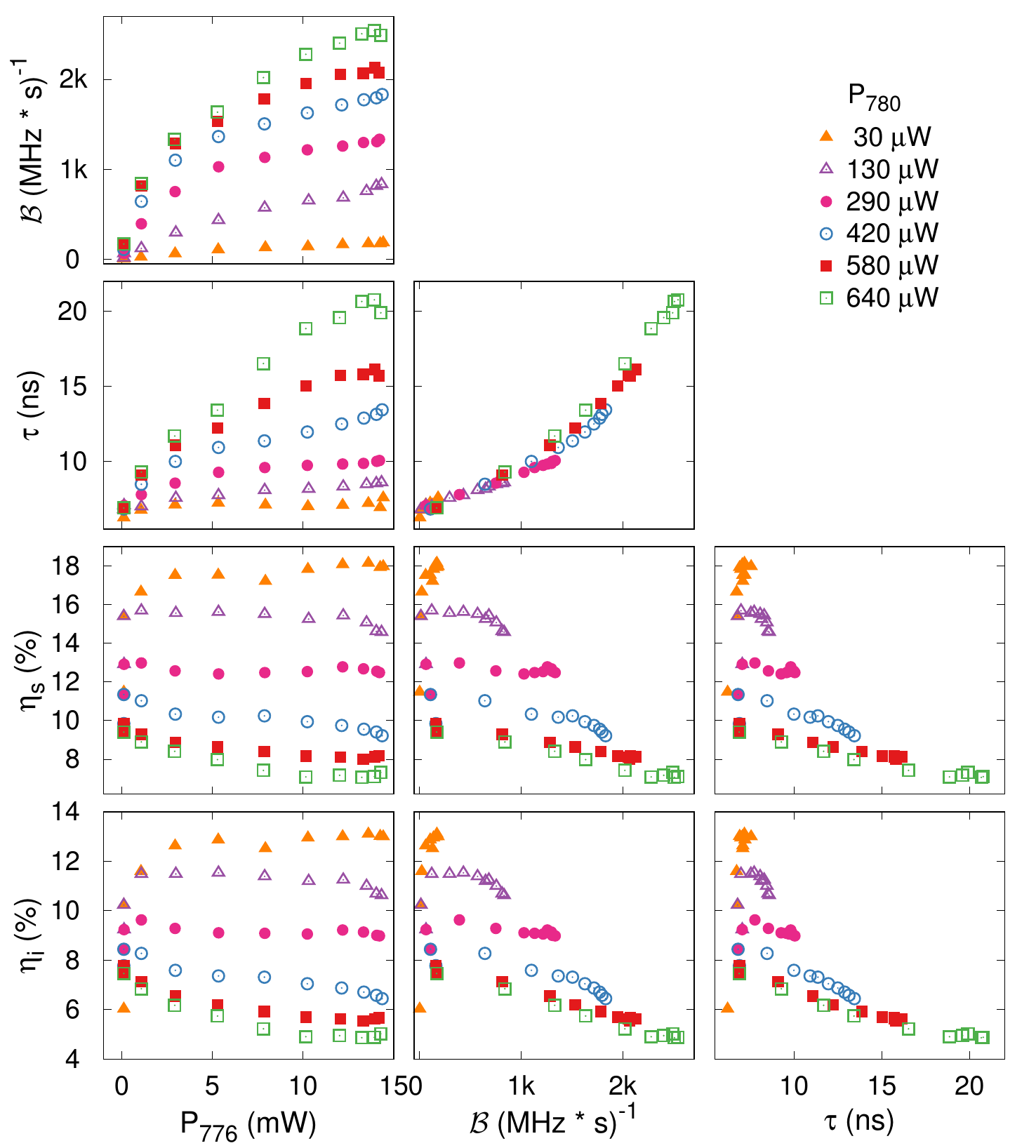}
      \caption{\label{fig:power_table}
    Summary of the effect of pump powers~$P_1$ and~$P_1$
    on heralding efficiencies~$\eta_{s, i}$, coherence time~$\tau$, and spectral brightness~$\mathcal{B}$.
      Other parameters:
      OD=29,
      $\Delta=-60 $\,MHz,
      $\delta=3$\,MHz.
      }
  \end{figure}

  \begin{figure}
    \centering
      \includegraphics[width=\columnwidth]{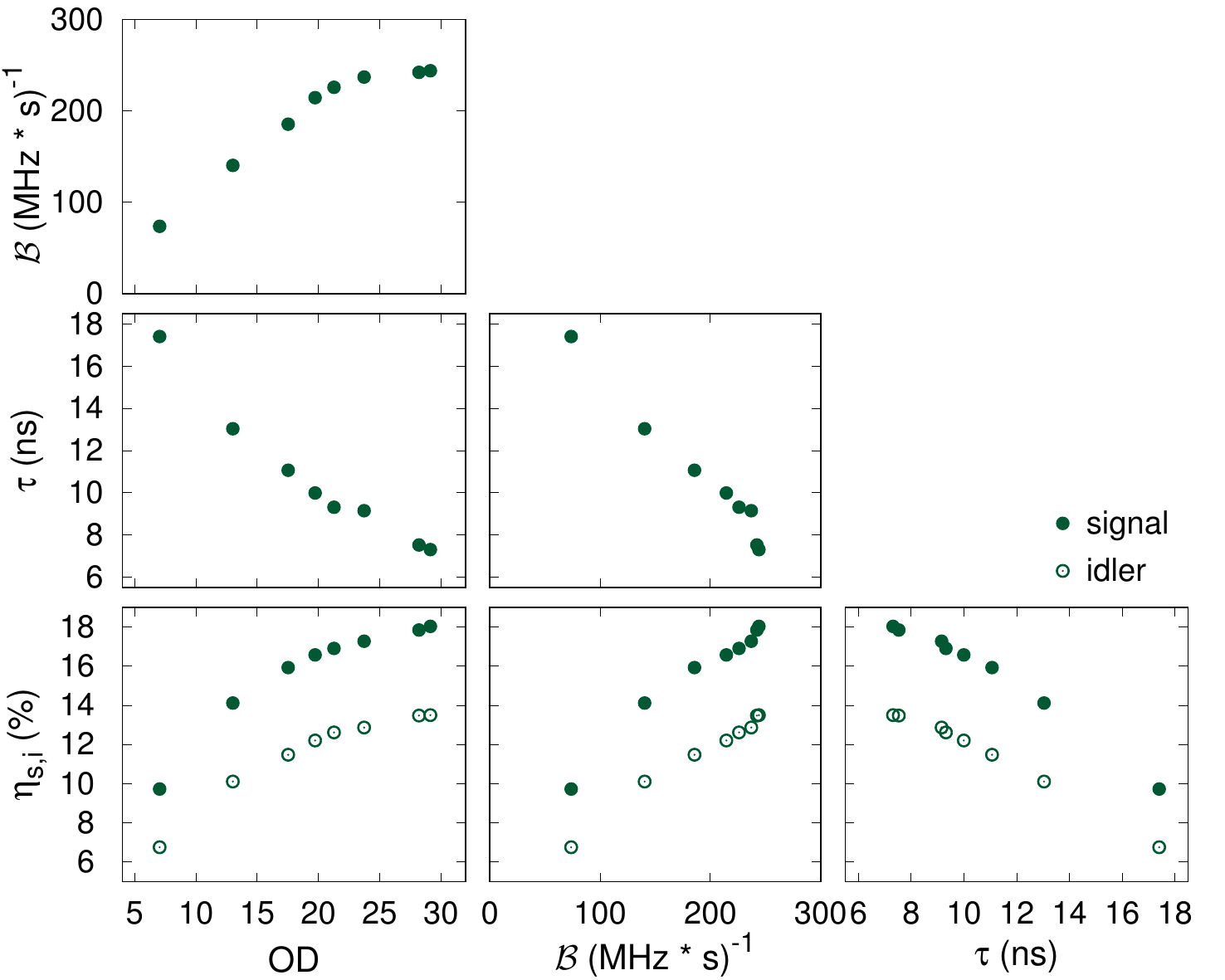}
      \caption{\label{fig:od_table}
    Summary of the effect of optical density~OD
    on heralding efficiencies~$\eta_{s, i}$, coherence time~$\tau$, and spectral brightness~$\mathcal{B}$.
      Other parameters:
      \mbox{$P_{776} = 15$\,mW},
      \mbox{$P_{780} = 300$\,$\mu$W},
      \mbox{$\Delta=-60$\,MHz},
      \mbox{$\delta=12$\,MHz}.
      }
  \end{figure}
  Following our characterization of this photon pair source,
  it is useful to introduce some guidelines for the choice of operational parameters.
  We summarize the effects of the different experimental knobs
  in Fig.~\ref{fig:detuning_table}, \ref{fig:power_table}, and~\ref{fig:od_table}.
  We included the heralding efficiency, coherence times, and spectral brightness~$\mathcal{B} = 2\pi\cdot\tau\cdot r_p$.
  Some trends are common: heralding efficiencies and coherence time appear to be inversely correlated, independently of the parameters we are varying.
  In experiments where the generated photon pairs interact with atomic systems it is often important to maximize the spectral brightness.
  In this case, it is necessary to maximize the optical density, set the two-photon detuning a few MHz red off resonance, and maximize both pump powers.
  If the target is to maximize the heralding efficiency, it is convenient to
  increase the two-photon detuning, and reduce power $P_{780}$ until a
  suitable compromise between heralding efficiency and brightness is reached.

\section{Conclusion}
  We presented an experimental study of the effect of two-photon detuning, pump intensity, and number of atoms on the generation rates and bandwidth of photon pairs from four-wave mixing in a cold ensemble of rubidium atoms.
  The study is useful to understand how to set the different parameters to better exploit the source characteristics, in particular when combined with other, generally very demanding, atomic systems~\cite{Leong:2015eb,Leong:2016jb}.

The effect of pump powers and two-photons detuning
on pair rates and efficiencies are compatible with the theoretical model presented by Whitley and Stroud~\cite{Whitley:1976ej}.
  An increase in pump power corresponds to an increase of pair and singles
  rates until a saturation level, with heralding efficiency determined mostly
  by the ground-state resonant pump. We can also explain the connection
  between the coincidence to accidentals ratio (CAR) and the generated pair
  rates.
  All rates increase with a reduction of the two-photons detuning at the expenses of heralding efficiency. This is well captured by the model, and can be intuitively explained as the result of competition between coherent and incoherent scatting processes excited by the same optical pumps.

  One of the attractive aspects of cold-atoms based photon pairs sources is their frequency characteristics: the generated pairs are usually resonant or close to resonant with their bandwidth of the same order of magnitude as atomic transitions.
  In our source the central wavelengths are fixed, the bandwidth instead is a function of the experimental parameters, in particular of the number of atoms.
  The dipole-dipole interaction between atoms gives rise to superradiance~\cite{Dicke:1954}, as evidenced by the reduction of coherence time as the number of atoms increases~\cite{Jen:2012}.
  But the total number of atoms is also a function of duration, intensity, and detuning of the pump beams because of optical pumping.
  The dynamics of the combined effect of collective interaction between atoms and optical pumping increases the complexity of the phenomenon,
  and
  we currently do not have a model that fully explain our result.
  Nonetheless, the experimental measurements are a useful guide to choose the number of atoms, together with the other parameters, that optimizes the specific properties desired from the source: rate, heralding efficiency, or bandwidth.

\section{Acknowledgments}
We like to thank Mathias A. Seidler, Matthias Steiner, and Chin Yue Sum for useful discussions about the theoretical modeling of the source.
This work was supported by the Ministry of Education in Singapore and the
National Research Foundation, Prime Minister’s office (partly under grant no
NRF-CRP12-2013-03).

\bibliographystyle{apsrev4-1}

\begin{thebibliography}{37}%
\makeatletter
\providecommand \@ifxundefined [1]{%
 \@ifx{#1\undefined}
}%
\providecommand \@ifnum [1]{%
 \ifnum #1\expandafter \@firstoftwo
 \else \expandafter \@secondoftwo
 \fi
}%
\providecommand \@ifx [1]{%
 \ifx #1\expandafter \@firstoftwo
 \else \expandafter \@secondoftwo
 \fi
}%
\providecommand \natexlab [1]{#1}%
\providecommand \enquote  [1]{``#1''}%
\providecommand \bibnamefont  [1]{#1}%
\providecommand \bibfnamefont [1]{#1}%
\providecommand \citenamefont [1]{#1}%
\providecommand \href@noop [0]{\@secondoftwo}%
\providecommand \href [0]{\begingroup \@sanitize@url \@href}%
\providecommand \@href[1]{\@@startlink{#1}\@@href}%
\providecommand \@@href[1]{\endgroup#1\@@endlink}%
\providecommand \@sanitize@url [0]{\catcode `\\12\catcode `\$12\catcode
  `\&12\catcode `\#12\catcode `\^12\catcode `\_12\catcode `\%12\relax}%
\providecommand \@@startlink[1]{}%
\providecommand \@@endlink[0]{}%
\providecommand \url  [0]{\begingroup\@sanitize@url \@url }%
\providecommand \@url [1]{\endgroup\@href {#1}{\urlprefix }}%
\providecommand \urlprefix  [0]{URL }%
\providecommand \Eprint [0]{\href }%
\providecommand \doibase [0]{http://dx.doi.org/}%
\providecommand \selectlanguage [0]{\@gobble}%
\providecommand \bibinfo  [0]{\@secondoftwo}%
\providecommand \bibfield  [0]{\@secondoftwo}%
\providecommand \translation [1]{[#1]}%
\providecommand \BibitemOpen [0]{}%
\providecommand \bibitemStop [0]{}%
\providecommand \bibitemNoStop [0]{.\EOS\space}%
\providecommand \EOS [0]{\spacefactor3000\relax}%
\providecommand \BibitemShut  [1]{\csname bibitem#1\endcsname}%
\let\auto@bib@innerbib\@empty
\bibitem [{\citenamefont {Clauser}\ and\ \citenamefont
  {Shimony}(1978)}]{Clauser:1978jf}%
  \BibitemOpen
  \bibfield  {author} {\bibinfo {author} {\bibfnamefont {J.~F.}\ \bibnamefont
  {Clauser}}\ and\ \bibinfo {author} {\bibfnamefont {A.}~\bibnamefont
  {Shimony}},\ }\href {\doibase 10.1088/0034-4885/41/12/002} {\bibfield
  {journal} {\bibinfo  {journal} {Rept.Prog.Phys.}\ }\textbf {\bibinfo {volume}
  {41}},\ \bibinfo {pages} {1881} (\bibinfo {year} {1978})}\BibitemShut
  {NoStop}%
\bibitem [{\citenamefont {Aspect}\ \emph {et~al.}(1981)\citenamefont {Aspect},
  \citenamefont {Grangier},\ and\ \citenamefont {Roger}}]{Aspect:1981ga}%
  \BibitemOpen
  \bibfield  {author} {\bibinfo {author} {\bibfnamefont {A.}~\bibnamefont
  {Aspect}}, \bibinfo {author} {\bibfnamefont {P.}~\bibnamefont {Grangier}}, \
  and\ \bibinfo {author} {\bibfnamefont {G.}~\bibnamefont {Roger}},\ }\href
  {\doibase 10.1103/PhysRevLett.47.460} {\bibfield  {journal} {\bibinfo
  {journal} {Phys. Rev. Lett.}\ }\textbf {\bibinfo {volume} {47}},\ \bibinfo
  {pages} {460} (\bibinfo {year} {1981})}\BibitemShut {NoStop}%
\bibitem [{\citenamefont {Ekert}(1991)}]{Ekert:1991zz}%
  \BibitemOpen
  \bibfield  {author} {\bibinfo {author} {\bibfnamefont {A.~K.}\ \bibnamefont
  {Ekert}},\ }\href {\doibase 10.1103/PhysRevLett.67.661} {\bibfield  {journal}
  {\bibinfo  {journal} {Phys. Rev. Lett.}\ }\textbf {\bibinfo {volume} {67}},\
  \bibinfo {pages} {661} (\bibinfo {year} {1991})}\BibitemShut {NoStop}%
\bibitem [{\citenamefont {Bouwmeester}\ \emph {et~al.}(1997)\citenamefont
  {Bouwmeester}, \citenamefont {Pan}, \citenamefont {Mattle}, \citenamefont
  {Daniell}, \citenamefont {Zeilinger},\ and\ \citenamefont
  {Weinfurter}}]{Bouwmeester:1997wk}%
  \BibitemOpen
  \bibfield  {author} {\bibinfo {author} {\bibfnamefont {D.}~\bibnamefont
  {Bouwmeester}}, \bibinfo {author} {\bibfnamefont {J.-W.}\ \bibnamefont
  {Pan}}, \bibinfo {author} {\bibfnamefont {K.}~\bibnamefont {Mattle}},
  \bibinfo {author} {\bibfnamefont {M.}~\bibnamefont {Daniell}}, \bibinfo
  {author} {\bibfnamefont {A.}~\bibnamefont {Zeilinger}}, \ and\ \bibinfo
  {author} {\bibfnamefont {H.}~\bibnamefont {Weinfurter}},\ }\href
  {http://www.nature.com/nature/journal/v390/n6660/abs/390575a0.html}
  {\bibfield  {journal} {\bibinfo  {journal} {Nature}\ }\textbf {\bibinfo
  {volume} {390}},\ \bibinfo {pages} {575} (\bibinfo {year}
  {1997})}\BibitemShut {NoStop}%
\bibitem [{\citenamefont {Boschi}\ \emph {et~al.}(1998)\citenamefont {Boschi},
  \citenamefont {Branca}, \citenamefont {De~Martini}, \citenamefont {Hardy},\
  and\ \citenamefont {Popescu}}]{Boschi:1998iu}%
  \BibitemOpen
  \bibfield  {author} {\bibinfo {author} {\bibfnamefont {D.}~\bibnamefont
  {Boschi}}, \bibinfo {author} {\bibfnamefont {S.}~\bibnamefont {Branca}},
  \bibinfo {author} {\bibfnamefont {F.}~\bibnamefont {De~Martini}}, \bibinfo
  {author} {\bibfnamefont {L.}~\bibnamefont {Hardy}}, \ and\ \bibinfo {author}
  {\bibfnamefont {S.}~\bibnamefont {Popescu}},\ }\href {\doibase
  10.1103/PhysRevLett.80.1121} {\bibfield  {journal} {\bibinfo  {journal}
  {Phys. Rev. Lett.}\ }\textbf {\bibinfo {volume} {301}},\ \bibinfo {pages}
  {1121} (\bibinfo {year} {1998})}\BibitemShut {NoStop}%
\bibitem [{\citenamefont {Burnham}\ and\ \citenamefont
  {Weinberg}(1970)}]{Burnham:1970gz}%
  \BibitemOpen
  \bibfield  {author} {\bibinfo {author} {\bibfnamefont {D.~C.}\ \bibnamefont
  {Burnham}}\ and\ \bibinfo {author} {\bibfnamefont {D.~L.}\ \bibnamefont
  {Weinberg}},\ }\href {\doibase 10.1103/PhysRevLett.25.84} {\bibfield
  {journal} {\bibinfo  {journal} {Phys. Rev. Lett.}\ }\textbf {\bibinfo
  {volume} {25}},\ \bibinfo {pages} {84} (\bibinfo {year} {1970})}\BibitemShut
  {NoStop}%
\bibitem [{\citenamefont {Kwiat}\ \emph {et~al.}(1995)\citenamefont {Kwiat},
  \citenamefont {Mattle}, \citenamefont {Weinfurter}, \citenamefont
  {Zeilinger}, \citenamefont {Sergienko},\ and\ \citenamefont
  {Shih}}]{Kwiat:1995ub}%
  \BibitemOpen
  \bibfield  {author} {\bibinfo {author} {\bibfnamefont {P.~G.}\ \bibnamefont
  {Kwiat}}, \bibinfo {author} {\bibfnamefont {K.}~\bibnamefont {Mattle}},
  \bibinfo {author} {\bibfnamefont {H.}~\bibnamefont {Weinfurter}}, \bibinfo
  {author} {\bibfnamefont {A.}~\bibnamefont {Zeilinger}}, \bibinfo {author}
  {\bibfnamefont {A.~V.}\ \bibnamefont {Sergienko}}, \ and\ \bibinfo {author}
  {\bibfnamefont {Y.}~\bibnamefont {Shih}},\ }\href
  {http://link.aps.org/doi/10.1103/PhysRevLett.75.4337} {\bibfield  {journal}
  {\bibinfo  {journal} {Phys. Rev. Lett.}\ }\textbf {\bibinfo {volume} {75}},\
  \bibinfo {pages} {4337} (\bibinfo {year} {1995})}\BibitemShut {NoStop}%
\bibitem [{\citenamefont {Kurtsiefer}\ \emph {et~al.}(2001)\citenamefont
  {Kurtsiefer}, \citenamefont {Oberparleiter},\ and\ \citenamefont
  {Weinfurter}}]{Kurtsiefer:2001ha}%
  \BibitemOpen
  \bibfield  {author} {\bibinfo {author} {\bibfnamefont {C.}~\bibnamefont
  {Kurtsiefer}}, \bibinfo {author} {\bibfnamefont {M.}~\bibnamefont
  {Oberparleiter}}, \ and\ \bibinfo {author} {\bibfnamefont {H.}~\bibnamefont
  {Weinfurter}},\ }\href {\doibase 10.1103/PhysRevA.64.023802} {\bibfield
  {journal} {\bibinfo  {journal} {Phys. Rev. A}\ }\textbf {\bibinfo {volume}
  {64}},\ \bibinfo {pages} {023802} (\bibinfo {year} {2001})}\BibitemShut
  {NoStop}%
\bibitem [{\citenamefont {Kuklewicz}\ \emph {et~al.}(2006)\citenamefont
  {Kuklewicz}, \citenamefont {Wong},\ and\ \citenamefont
  {Shapiro}}]{Kuklewicz:2006gq}%
  \BibitemOpen
  \bibfield  {author} {\bibinfo {author} {\bibfnamefont {C.~E.}\ \bibnamefont
  {Kuklewicz}}, \bibinfo {author} {\bibfnamefont {F.~N.~C.}\ \bibnamefont
  {Wong}}, \ and\ \bibinfo {author} {\bibfnamefont {J.~H.}\ \bibnamefont
  {Shapiro}},\ }\href {\doibase 10.1103/PhysRevLett.97.223601} {\bibfield
  {journal} {\bibinfo  {journal} {Phys. Rev. Lett.}\ }\textbf {\bibinfo
  {volume} {97}},\ \bibinfo {pages} {223601} (\bibinfo {year}
  {2006})}\BibitemShut {NoStop}%
\bibitem [{\citenamefont {Wolfgramm}\ \emph {et~al.}(2008)\citenamefont
  {Wolfgramm}, \citenamefont {Xing}, \citenamefont {Cer{\`e}}, \citenamefont
  {Predojevi{\'c}}, \citenamefont {Steinberg},\ and\ \citenamefont
  {Mitchell}}]{Wolfgramm:08}%
  \BibitemOpen
  \bibfield  {author} {\bibinfo {author} {\bibfnamefont {F.}~\bibnamefont
  {Wolfgramm}}, \bibinfo {author} {\bibfnamefont {X.}~\bibnamefont {Xing}},
  \bibinfo {author} {\bibfnamefont {A.}~\bibnamefont {Cer{\`e}}}, \bibinfo
  {author} {\bibfnamefont {A.}~\bibnamefont {Predojevi{\'c}}}, \bibinfo
  {author} {\bibfnamefont {A.~M.}\ \bibnamefont {Steinberg}}, \ and\ \bibinfo
  {author} {\bibfnamefont {M.~W.}\ \bibnamefont {Mitchell}},\ }\href {\doibase
  10.1364/OE.16.018145} {\bibfield  {journal} {\bibinfo  {journal} {Opt.
  Express}\ }\textbf {\bibinfo {volume} {16}},\ \bibinfo {pages} {18145}
  (\bibinfo {year} {2008})}\BibitemShut {NoStop}%
\bibitem [{\citenamefont {Fekete}\ \emph {et~al.}(2013)\citenamefont {Fekete},
  \citenamefont {Rielander}, \citenamefont {Cristiani},\ and\ \citenamefont
  {de~Riedmatten}}]{Fekete:2013kr}%
  \BibitemOpen
  \bibfield  {author} {\bibinfo {author} {\bibfnamefont {J.}~\bibnamefont
  {Fekete}}, \bibinfo {author} {\bibfnamefont {D.}~\bibnamefont {Rielander}},
  \bibinfo {author} {\bibfnamefont {M.}~\bibnamefont {Cristiani}}, \ and\
  \bibinfo {author} {\bibfnamefont {H.}~\bibnamefont {de~Riedmatten}},\ }\href
  {\doibase 10.1103/PhysRevLett.110.220502} {\bibfield  {journal} {\bibinfo
  {journal} {Phys. Rev. Lett.}\ }\textbf {\bibinfo {volume} {110}},\ \bibinfo
  {pages} {220502} (\bibinfo {year} {2013})}\BibitemShut {NoStop}%
\bibitem [{\citenamefont {Neergaard-Nielsen}\ \emph {et~al.}(2007)\citenamefont
  {Neergaard-Nielsen}, \citenamefont {Nielsen}, \citenamefont {Nielsen},
  \citenamefont {Takahashi}, \citenamefont {Vistnes},\ and\ \citenamefont
  {Polzik}}]{NeergaardNielsen:2007wa}%
  \BibitemOpen
  \bibfield  {author} {\bibinfo {author} {\bibfnamefont {J.~S.}\ \bibnamefont
  {Neergaard-Nielsen}}, \bibinfo {author} {\bibfnamefont {B.~M.}\ \bibnamefont
  {Nielsen}}, \bibinfo {author} {\bibfnamefont {B.~M.}\ \bibnamefont
  {Nielsen}}, \bibinfo {author} {\bibfnamefont {H.}~\bibnamefont {Takahashi}},
  \bibinfo {author} {\bibfnamefont {A.~I.}\ \bibnamefont {Vistnes}}, \ and\
  \bibinfo {author} {\bibfnamefont {E.~S.}\ \bibnamefont {Polzik}},\ }\href
  {\doibase 10.1364/OE.15.007940} {\bibfield  {journal} {\bibinfo  {journal}
  {Opt. Express}\ }\textbf {\bibinfo {volume} {15}},\ \bibinfo {pages} {7940}
  (\bibinfo {year} {2007})}\BibitemShut {NoStop}%
\bibitem [{\citenamefont {Haase}\ \emph {et~al.}(2009)\citenamefont {Haase},
  \citenamefont {Piro}, \citenamefont {Eschner},\ and\ \citenamefont
  {Mitchell}}]{Haase:2009ez}%
  \BibitemOpen
  \bibfield  {author} {\bibinfo {author} {\bibfnamefont {A.}~\bibnamefont
  {Haase}}, \bibinfo {author} {\bibfnamefont {N.}~\bibnamefont {Piro}},
  \bibinfo {author} {\bibfnamefont {J.}~\bibnamefont {Eschner}}, \ and\
  \bibinfo {author} {\bibfnamefont {M.~W.}\ \bibnamefont {Mitchell}},\ }\href
  {\doibase 10.1364/OL.34.000055} {\bibfield  {journal} {\bibinfo  {journal}
  {Opt. Lett.}\ }\textbf {\bibinfo {volume} {34}},\ \bibinfo {pages} {55}
  (\bibinfo {year} {2009})}\BibitemShut {NoStop}%
\bibitem [{\citenamefont {Schunk}\ \emph {et~al.}(2016)\citenamefont {Schunk},
  \citenamefont {Vogl}, \citenamefont {Sedlmeir}, \citenamefont {Strekalov},
  \citenamefont {Otterpohl}, \citenamefont {Averchenko}, \citenamefont
  {Schwefel}, \citenamefont {Leuchs},\ and\ \citenamefont
  {Marquardt}}]{Schunk:2016fl}%
  \BibitemOpen
  \bibfield  {author} {\bibinfo {author} {\bibfnamefont {G.}~\bibnamefont
  {Schunk}}, \bibinfo {author} {\bibfnamefont {U.}~\bibnamefont {Vogl}},
  \bibinfo {author} {\bibfnamefont {F.}~\bibnamefont {Sedlmeir}}, \bibinfo
  {author} {\bibfnamefont {D.~V.}\ \bibnamefont {Strekalov}}, \bibinfo {author}
  {\bibfnamefont {A.}~\bibnamefont {Otterpohl}}, \bibinfo {author}
  {\bibfnamefont {V.}~\bibnamefont {Averchenko}}, \bibinfo {author}
  {\bibfnamefont {H.~G.~L.}\ \bibnamefont {Schwefel}}, \bibinfo {author}
  {\bibfnamefont {G.}~\bibnamefont {Leuchs}}, \ and\ \bibinfo {author}
  {\bibfnamefont {C.}~\bibnamefont {Marquardt}},\ }\href {\doibase
  10.1080/09500340.2016.1148211} {\bibfield  {journal} {\bibinfo  {journal}
  {Journal of Modern Optics}\ ,\ \bibinfo {pages} {1}} (\bibinfo {year}
  {2016})}\BibitemShut {NoStop}%
\bibitem [{\citenamefont {Braje}\ \emph {et~al.}(2004)\citenamefont {Braje},
  \citenamefont {Bali{\'c}}, \citenamefont {Goda}, \citenamefont {Yin},\ and\
  \citenamefont {Harris}}]{Braje:2004dt}%
  \BibitemOpen
  \bibfield  {author} {\bibinfo {author} {\bibfnamefont {D.~A.}\ \bibnamefont
  {Braje}}, \bibinfo {author} {\bibfnamefont {V.}~\bibnamefont {Bali{\'c}}},
  \bibinfo {author} {\bibfnamefont {S.}~\bibnamefont {Goda}}, \bibinfo {author}
  {\bibfnamefont {G.-Y.}\ \bibnamefont {Yin}}, \ and\ \bibinfo {author}
  {\bibfnamefont {S.~E.}\ \bibnamefont {Harris}},\ }\href {\doibase
  10.1103/PhysRevLett.93.183601} {\bibfield  {journal} {\bibinfo  {journal}
  {Phys. Rev. Lett.}\ }\textbf {\bibinfo {volume} {93}},\ \bibinfo {pages}
  {183601} (\bibinfo {year} {2004})}\BibitemShut {NoStop}%
\bibitem [{\citenamefont {Matsukevich}\ \emph {et~al.}(2005)\citenamefont
  {Matsukevich}, \citenamefont {Bhattacharya}, \citenamefont {Lan},
  \citenamefont {Jenkins},\ and\ \citenamefont {Kuzmich}}]{Matsukevich:2005fz}%
  \BibitemOpen
  \bibfield  {author} {\bibinfo {author} {\bibfnamefont {D.~N.}\ \bibnamefont
  {Matsukevich}}, \bibinfo {author} {\bibfnamefont {M.}~\bibnamefont
  {Bhattacharya}}, \bibinfo {author} {\bibfnamefont {S.~Y.}\ \bibnamefont
  {Lan}}, \bibinfo {author} {\bibfnamefont {S.~D.}\ \bibnamefont {Jenkins}}, \
  and\ \bibinfo {author} {\bibfnamefont {A.}~\bibnamefont {Kuzmich}},\ }\href
  {\doibase 10.1103/PhysRevLett.95.040405} {\bibfield  {journal} {\bibinfo
  {journal} {Phys. Rev. Lett.}\ }\textbf {\bibinfo {volume} {95}},\ \bibinfo
  {pages} {040405} (\bibinfo {year} {2005})}\BibitemShut {NoStop}%
\bibitem [{\citenamefont {Chen}\ and\ \citenamefont {Du}(2012)}]{Chen:2012jv}%
  \BibitemOpen
  \bibfield  {author} {\bibinfo {author} {\bibfnamefont {J.~F.}\ \bibnamefont
  {Chen}}\ and\ \bibinfo {author} {\bibfnamefont {S.}~\bibnamefont {Du}},\
  }\href {\doibase 10.1007/s11467-012-0260-1} {\bibfield  {journal} {\bibinfo
  {journal} {Front. Phys.}\ }\textbf {\bibinfo {volume} {7}},\ \bibinfo {pages}
  {494} (\bibinfo {year} {2012})}\BibitemShut {NoStop}%
\bibitem [{\citenamefont {Srivathsan}\ \emph {et~al.}(2013)\citenamefont
  {Srivathsan}, \citenamefont {Gulati}, \citenamefont {Chng}, \citenamefont
  {Maslennikov}, \citenamefont {Matsukevich},\ and\ \citenamefont
  {Kurtsiefer}}]{Srivathsan:2013fa}%
  \BibitemOpen
  \bibfield  {author} {\bibinfo {author} {\bibfnamefont {B.}~\bibnamefont
  {Srivathsan}}, \bibinfo {author} {\bibfnamefont {G.~K.}\ \bibnamefont
  {Gulati}}, \bibinfo {author} {\bibfnamefont {B.}~\bibnamefont {Chng}},
  \bibinfo {author} {\bibfnamefont {G.}~\bibnamefont {Maslennikov}}, \bibinfo
  {author} {\bibfnamefont {D.~N.}\ \bibnamefont {Matsukevich}}, \ and\ \bibinfo
  {author} {\bibfnamefont {C.}~\bibnamefont {Kurtsiefer}},\ }\href {\doibase
  10.1103/PhysRevLett.111.123602} {\bibfield  {journal} {\bibinfo  {journal}
  {Phys. Rev. Lett.}\ }\textbf {\bibinfo {volume} {111}},\ \bibinfo {pages}
  {123602} (\bibinfo {year} {2013})}\BibitemShut {NoStop}%
\bibitem [{\citenamefont {Gulati}\ \emph {et~al.}(2014)\citenamefont {Gulati},
  \citenamefont {Srivathsan}, \citenamefont {Chng}, \citenamefont {Cer\`e},
  \citenamefont {Matsukevich},\ and\ \citenamefont {Kurtsiefer}}]{GK_2014}%
  \BibitemOpen
  \bibfield  {author} {\bibinfo {author} {\bibfnamefont {G.~K.}\ \bibnamefont
  {Gulati}}, \bibinfo {author} {\bibfnamefont {B.}~\bibnamefont {Srivathsan}},
  \bibinfo {author} {\bibfnamefont {B.}~\bibnamefont {Chng}}, \bibinfo {author}
  {\bibfnamefont {A.}~\bibnamefont {Cer\`e}}, \bibinfo {author} {\bibfnamefont
  {D.}~\bibnamefont {Matsukevich}}, \ and\ \bibinfo {author} {\bibfnamefont
  {C.}~\bibnamefont {Kurtsiefer}},\ }\href {\doibase
  10.1103/PhysRevA.90.033819} {\bibfield  {journal} {\bibinfo  {journal} {Phys.
  Rev. A}\ }\textbf {\bibinfo {volume} {90}},\ \bibinfo {pages} {033819}
  (\bibinfo {year} {2014})}\BibitemShut {NoStop}%
\bibitem [{\citenamefont {Srivathsan}\ \emph {et~al.}(2014)\citenamefont
  {Srivathsan}, \citenamefont {Gulati}, \citenamefont {Cer{\`e}}, \citenamefont
  {Chng},\ and\ \citenamefont {Kurtsiefer}}]{Srivathsan:2014jx}%
  \BibitemOpen
  \bibfield  {author} {\bibinfo {author} {\bibfnamefont {B.}~\bibnamefont
  {Srivathsan}}, \bibinfo {author} {\bibfnamefont {G.~K.}\ \bibnamefont
  {Gulati}}, \bibinfo {author} {\bibfnamefont {A.}~\bibnamefont {Cer{\`e}}},
  \bibinfo {author} {\bibfnamefont {B.}~\bibnamefont {Chng}}, \ and\ \bibinfo
  {author} {\bibfnamefont {C.}~\bibnamefont {Kurtsiefer}},\ }\href {\doibase
  10.1103/PhysRevLett.113.163601} {\bibfield  {journal} {\bibinfo  {journal}
  {Phys. Rev. Lett.}\ }\textbf {\bibinfo {volume} {113}},\ \bibinfo {pages}
  {163601} (\bibinfo {year} {2014})}\BibitemShut {NoStop}%
\bibitem [{\citenamefont {Gulati}\ \emph {et~al.}(2015)\citenamefont {Gulati},
  \citenamefont {Srivathsan}, \citenamefont {Chng}, \citenamefont {Cer{\`e}},\
  and\ \citenamefont {Kurtsiefer}}]{Gulati:2015ee}%
  \BibitemOpen
  \bibfield  {author} {\bibinfo {author} {\bibfnamefont {G.~K.}\ \bibnamefont
  {Gulati}}, \bibinfo {author} {\bibfnamefont {B.}~\bibnamefont {Srivathsan}},
  \bibinfo {author} {\bibfnamefont {B.}~\bibnamefont {Chng}}, \bibinfo {author}
  {\bibfnamefont {A.}~\bibnamefont {Cer{\`e}}}, \ and\ \bibinfo {author}
  {\bibfnamefont {C.}~\bibnamefont {Kurtsiefer}},\ }\href {\doibase
  10.1088/1367-2630/17/9/093034} {\bibfield  {journal} {\bibinfo  {journal}
  {New J. Phys.}\ }\textbf {\bibinfo {volume} {17}},\ \bibinfo {pages} {093034}
  (\bibinfo {year} {2015})}\BibitemShut {NoStop}%
\bibitem [{\citenamefont {Leong}\ \emph {et~al.}(2015)\citenamefont {Leong},
  \citenamefont {Kosen}, \citenamefont {Srivathsan}, \citenamefont {Gulati},
  \citenamefont {Cer{\`e}},\ and\ \citenamefont {Kurtsiefer}}]{Leong:2015eb}%
  \BibitemOpen
  \bibfield  {author} {\bibinfo {author} {\bibfnamefont {V.}~\bibnamefont
  {Leong}}, \bibinfo {author} {\bibfnamefont {S.}~\bibnamefont {Kosen}},
  \bibinfo {author} {\bibfnamefont {B.}~\bibnamefont {Srivathsan}}, \bibinfo
  {author} {\bibfnamefont {G.~K.}\ \bibnamefont {Gulati}}, \bibinfo {author}
  {\bibfnamefont {A.}~\bibnamefont {Cer{\`e}}}, \ and\ \bibinfo {author}
  {\bibfnamefont {C.}~\bibnamefont {Kurtsiefer}},\ }\href {\doibase
  10.1103/PhysRevA.91.063829} {\bibfield  {journal} {\bibinfo  {journal} {Phys.
  Rev. A}\ }\textbf {\bibinfo {volume} {91}},\ \bibinfo {pages} {063829}
  (\bibinfo {year} {2015})}\BibitemShut {NoStop}%
\bibitem [{\citenamefont {Leong}\ \emph {et~al.}(2016)\citenamefont {Leong},
  \citenamefont {Seidler}, \citenamefont {Steiner}, \citenamefont {Cer{\`e}},\
  and\ \citenamefont {Kurtsiefer}}]{Leong:2016jb}%
  \BibitemOpen
  \bibfield  {author} {\bibinfo {author} {\bibfnamefont {V.}~\bibnamefont
  {Leong}}, \bibinfo {author} {\bibfnamefont {M.~A.}\ \bibnamefont {Seidler}},
  \bibinfo {author} {\bibfnamefont {M.}~\bibnamefont {Steiner}}, \bibinfo
  {author} {\bibfnamefont {A.}~\bibnamefont {Cer{\`e}}}, \ and\ \bibinfo
  {author} {\bibfnamefont {C.}~\bibnamefont {Kurtsiefer}},\ }\href {\doibase
  10.1038/ncomms13716} {\bibfield  {journal} {\bibinfo  {journal} {Nat Comms}\
  }\textbf {\bibinfo {volume} {7}},\ \bibinfo {pages} {13716} (\bibinfo {year}
  {2016})}\BibitemShut {NoStop}%
\bibitem [{\citenamefont {Chaneli{\`e}re}\ \emph {et~al.}(2006)\citenamefont
  {Chaneli{\`e}re}, \citenamefont {Matsukevich},\ and\ \citenamefont
  {Jenkins}}]{Chaneliere:2006}%
  \BibitemOpen
  \bibfield  {author} {\bibinfo {author} {\bibfnamefont {T.}~\bibnamefont
  {Chaneli{\`e}re}}, \bibinfo {author} {\bibfnamefont {D.~N.}\ \bibnamefont
  {Matsukevich}}, \ and\ \bibinfo {author} {\bibfnamefont {S.~D.}\ \bibnamefont
  {Jenkins}},\ }\href {\doibase 10.1103/PhysRevLett.96.093604} {\bibfield
  {journal} {\bibinfo  {journal} {Phys. Rev. Lett.}\ }\textbf {\bibinfo
  {volume} {96}},\ \bibinfo {pages} {093604} (\bibinfo {year}
  {2006})}\BibitemShut {NoStop}%
\bibitem [{\citenamefont {Mollow}(1969)}]{Mollow_1969}%
  \BibitemOpen
  \bibfield  {author} {\bibinfo {author} {\bibfnamefont {B.~R.}\ \bibnamefont
  {Mollow}},\ }\href {\doibase 10.1103/PhysRev.188.1969} {\bibfield  {journal}
  {\bibinfo  {journal} {Phys. Rev.}\ }\textbf {\bibinfo {volume} {188}},\
  \bibinfo {pages} {1969} (\bibinfo {year} {1969})}\BibitemShut {NoStop}%
\bibitem [{\citenamefont {Cohen-Tannoudji}\ \emph {et~al.}(1994)\citenamefont
  {Cohen-Tannoudji}, \citenamefont {Dupont-Roc},\ and\ \citenamefont
  {Grynberg}}]{cohen:2008}%
  \BibitemOpen
  \bibfield  {author} {\bibinfo {author} {\bibfnamefont {C.}~\bibnamefont
  {Cohen-Tannoudji}}, \bibinfo {author} {\bibfnamefont {J.}~\bibnamefont
  {Dupont-Roc}}, \ and\ \bibinfo {author} {\bibfnamefont {G.}~\bibnamefont
  {Grynberg}},\ }\enquote {\bibinfo {title} {Optical bloch equations},}\ in\
  \href@noop {} {\emph {\bibinfo {booktitle} {Advances in Cryptology - Proc.
  Eurocrypt'94}}}\ (\bibinfo  {publisher} {Wiley-VCH Verlag GmbH},\ \bibinfo
  {year} {1994})\ pp.\ \bibinfo {pages} {353--405}\BibitemShut {NoStop}%
\bibitem [{\citenamefont {Whitley}\ and\ \citenamefont
  {C~R~Stroud}(1976)}]{Whitley:1976ej}%
  \BibitemOpen
  \bibfield  {author} {\bibinfo {author} {\bibfnamefont {R.~M.}\ \bibnamefont
  {Whitley}}\ and\ \bibinfo {author} {\bibfnamefont {J.}~\bibnamefont
  {C~R~Stroud}},\ }\href {\doibase 10.1103/PhysRevA.14.1498} {\bibfield
  {journal} {\bibinfo  {journal} {Phys. Rev. A}\ }\textbf {\bibinfo {volume}
  {14}},\ \bibinfo {pages} {1498} (\bibinfo {year} {1976})}\BibitemShut
  {NoStop}%
\bibitem [{\citenamefont {Lawande}\ \emph {et~al.}(1986)\citenamefont
  {Lawande}, \citenamefont {Puri},\ and\ \citenamefont
  {D{\textquoteright}Souza}}]{Lawande:1986ch}%
  \BibitemOpen
  \bibfield  {author} {\bibinfo {author} {\bibfnamefont {S.~V.}\ \bibnamefont
  {Lawande}}, \bibinfo {author} {\bibfnamefont {R.~R.}\ \bibnamefont {Puri}}, \
  and\ \bibinfo {author} {\bibfnamefont {R.}~\bibnamefont
  {D{\textquoteright}Souza}},\ }\href {\doibase 10.1103/PhysRevA.33.2504}
  {\bibfield  {journal} {\bibinfo  {journal} {Phys. Rev. A}\ }\textbf {\bibinfo
  {volume} {33}},\ \bibinfo {pages} {2504} (\bibinfo {year}
  {1986})}\BibitemShut {NoStop}%
\bibitem [{\citenamefont {McDonnell}\ \emph {et~al.}(2004)\citenamefont
  {McDonnell}, \citenamefont {Stacey},\ and\ \citenamefont
  {Steane}}]{McDonnell:2004db}%
  \BibitemOpen
  \bibfield  {author} {\bibinfo {author} {\bibfnamefont {M.~J.}\ \bibnamefont
  {McDonnell}}, \bibinfo {author} {\bibfnamefont {D.~N.}\ \bibnamefont
  {Stacey}}, \ and\ \bibinfo {author} {\bibfnamefont {A.~M.}\ \bibnamefont
  {Steane}},\ }\href {\doibase 10.1103/PhysRevA.70.053802} {\bibfield
  {journal} {\bibinfo  {journal} {Phys. Rev. A}\ }\textbf {\bibinfo {volume}
  {70}},\ \bibinfo {pages} {S123} (\bibinfo {year} {2004})}\BibitemShut
  {NoStop}%
\bibitem [{\citenamefont {Jen}(2012)}]{Jen:2012}%
  \BibitemOpen
  \bibfield  {author} {\bibinfo {author} {\bibfnamefont {H.~H.}\ \bibnamefont
  {Jen}},\ }\href {\doibase 10.1088/0953-4075/45/16/165504} {\bibfield
  {journal} {\bibinfo  {journal} {J. Phys. B: At. Mol. Opt. Phys.}\ }\textbf
  {\bibinfo {volume} {45}},\ \bibinfo {pages} {165504} (\bibinfo {year}
  {2012})}\BibitemShut {NoStop}%
\bibitem [{\citenamefont {Willis}\ \emph {et~al.}(2010)\citenamefont {Willis},
  \citenamefont {Becerra}, \citenamefont {Orozco},\ and\ \citenamefont
  {Rolston}}]{Willis:2010gp}%
  \BibitemOpen
  \bibfield  {author} {\bibinfo {author} {\bibfnamefont {R.~T.}\ \bibnamefont
  {Willis}}, \bibinfo {author} {\bibfnamefont {F.~E.}\ \bibnamefont {Becerra}},
  \bibinfo {author} {\bibfnamefont {L.~A.}\ \bibnamefont {Orozco}}, \ and\
  \bibinfo {author} {\bibfnamefont {S.~L.}\ \bibnamefont {Rolston}},\ }\href
  {\doibase 10.1103/PhysRevA.82.053842} {\bibfield  {journal} {\bibinfo
  {journal} {Phys. Rev. A}\ }\textbf {\bibinfo {volume} {82}},\ \bibinfo
  {pages} {053842} (\bibinfo {year} {2010})}\BibitemShut {NoStop}%
\bibitem [{\citenamefont {Ding}\ \emph {et~al.}(2012)\citenamefont {Ding},
  \citenamefont {Zhou}, \citenamefont {Shi}, \citenamefont {Zou},\ and\
  \citenamefont {Guo}}]{Ding:2012gh}%
  \BibitemOpen
  \bibfield  {author} {\bibinfo {author} {\bibfnamefont {D.-S.}\ \bibnamefont
  {Ding}}, \bibinfo {author} {\bibfnamefont {Z.-Y.}\ \bibnamefont {Zhou}},
  \bibinfo {author} {\bibfnamefont {B.-S.}\ \bibnamefont {Shi}}, \bibinfo
  {author} {\bibfnamefont {X.-B.}\ \bibnamefont {Zou}}, \ and\ \bibinfo
  {author} {\bibfnamefont {G.-C.}\ \bibnamefont {Guo}},\ }\href {\doibase
  10.1364/OE.20.011433} {\bibfield  {journal} {\bibinfo  {journal} {Opt.
  Express}\ }\textbf {\bibinfo {volume} {20}},\ \bibinfo {pages} {11433}
  (\bibinfo {year} {2012})}\BibitemShut {NoStop}%
\bibitem [{\citenamefont {Gross}\ and\ \citenamefont
  {Haroche}(1982)}]{Gross1982}%
  \BibitemOpen
  \bibfield  {author} {\bibinfo {author} {\bibfnamefont {M.}~\bibnamefont
  {Gross}}\ and\ \bibinfo {author} {\bibfnamefont {S.}~\bibnamefont
  {Haroche}},\ }\href {\doibase http://dx.doi.org/10.1016/0370-1573(82)90102-8}
  {\bibfield  {journal} {\bibinfo  {journal} {Physics Reports}\ }\textbf
  {\bibinfo {volume} {93}},\ \bibinfo {pages} {301 } (\bibinfo {year}
  {1982})}\BibitemShut {NoStop}%
\bibitem [{\citenamefont {Takesue}\ and\ \citenamefont
  {Shimizu}(2010)}]{Takesue:2010gd}%
  \BibitemOpen
  \bibfield  {author} {\bibinfo {author} {\bibfnamefont {H.}~\bibnamefont
  {Takesue}}\ and\ \bibinfo {author} {\bibfnamefont {K.}~\bibnamefont
  {Shimizu}},\ }\href {\doibase 10.1016/j.optcom.2009.10.008} {\bibfield
  {journal} {\bibinfo  {journal} {Opt. Commun.}\ }\textbf {\bibinfo {volume}
  {283}},\ \bibinfo {pages} {276} (\bibinfo {year} {2010})}\BibitemShut
  {NoStop}%
\bibitem [{\citenamefont {Xiong}\ \emph {et~al.}(2011)\citenamefont {Xiong},
  \citenamefont {Marshall}, \citenamefont {Peruzzo}, \citenamefont {Lobino},
  \citenamefont {Clark}, \citenamefont {Choi}, \citenamefont {Madden},
  \citenamefont {Natarajan}, \citenamefont {Tanner}, \citenamefont {Hadfield},
  \citenamefont {Dorenbos}, \citenamefont {Zijlstra}, \citenamefont {Zwiller},
  \citenamefont {Thompson}, \citenamefont {Rarity}, \citenamefont {Steel},
  \citenamefont {Luther-Davies}, \citenamefont {Eggleton},\ and\ \citenamefont
  {O{\textquoteright}Brien}}]{Xiong:2011kq}%
  \BibitemOpen
  \bibfield  {author} {\bibinfo {author} {\bibfnamefont {C.}~\bibnamefont
  {Xiong}}, \bibinfo {author} {\bibfnamefont {G.~D.}\ \bibnamefont {Marshall}},
  \bibinfo {author} {\bibfnamefont {A.}~\bibnamefont {Peruzzo}}, \bibinfo
  {author} {\bibfnamefont {M.}~\bibnamefont {Lobino}}, \bibinfo {author}
  {\bibfnamefont {A.~S.}\ \bibnamefont {Clark}}, \bibinfo {author}
  {\bibfnamefont {D.~Y.}\ \bibnamefont {Choi}}, \bibinfo {author}
  {\bibfnamefont {S.~J.}\ \bibnamefont {Madden}}, \bibinfo {author}
  {\bibfnamefont {C.~M.}\ \bibnamefont {Natarajan}}, \bibinfo {author}
  {\bibfnamefont {M.~G.}\ \bibnamefont {Tanner}}, \bibinfo {author}
  {\bibfnamefont {R.~H.}\ \bibnamefont {Hadfield}}, \bibinfo {author}
  {\bibfnamefont {S.~N.}\ \bibnamefont {Dorenbos}}, \bibinfo {author}
  {\bibfnamefont {T.}~\bibnamefont {Zijlstra}}, \bibinfo {author}
  {\bibfnamefont {V.}~\bibnamefont {Zwiller}}, \bibinfo {author} {\bibfnamefont
  {M.~G.}\ \bibnamefont {Thompson}}, \bibinfo {author} {\bibfnamefont {J.~G.}\
  \bibnamefont {Rarity}}, \bibinfo {author} {\bibfnamefont {M.~J.}\
  \bibnamefont {Steel}}, \bibinfo {author} {\bibfnamefont {B.}~\bibnamefont
  {Luther-Davies}}, \bibinfo {author} {\bibfnamefont {B.~J.}\ \bibnamefont
  {Eggleton}}, \ and\ \bibinfo {author} {\bibfnamefont {J.~L.}\ \bibnamefont
  {O{\textquoteright}Brien}},\ }\href {\doibase 10.1063/1.3549744} {\bibfield
  {journal} {\bibinfo  {journal} {Appl. Phys. Lett.}\ }\textbf {\bibinfo
  {volume} {98}},\ \bibinfo {pages} {051101} (\bibinfo {year}
  {2011})}\BibitemShut {NoStop}%
\bibitem [{\citenamefont {Dicke}(1954)}]{Dicke:1954}%
  \BibitemOpen
  \bibfield  {author} {\bibinfo {author} {\bibfnamefont {R.~H.}\ \bibnamefont
  {Dicke}},\ }\href@noop {} {\bibfield  {journal} {\bibinfo  {journal} {Phys.
  Rev.}\ }\textbf {\bibinfo {volume} {93}},\ \bibinfo {pages} {99} (\bibinfo
  {year} {1954})}\BibitemShut {NoStop}%
\bibitem [{\citenamefont {Akimov}\ \emph {et~al.}(2010)\citenamefont {Akimov},
  \citenamefont {Tereshchenko}, \citenamefont {Snigirev}, \citenamefont
  {Samokotin}, \citenamefont {Sokolov},\ and\ \citenamefont
  {Sorokin}}]{Akimov:2010hm}%
  \BibitemOpen
  \bibfield  {author} {\bibinfo {author} {\bibfnamefont {A.~V.}\ \bibnamefont
  {Akimov}}, \bibinfo {author} {\bibfnamefont {E.~O.}\ \bibnamefont
  {Tereshchenko}}, \bibinfo {author} {\bibfnamefont {S.~A.}\ \bibnamefont
  {Snigirev}}, \bibinfo {author} {\bibfnamefont {A.~Y.}\ \bibnamefont
  {Samokotin}}, \bibinfo {author} {\bibfnamefont {A.~V.}\ \bibnamefont
  {Sokolov}}, \ and\ \bibinfo {author} {\bibfnamefont {V.~N.}\ \bibnamefont
  {Sorokin}},\ }\href {\doibase 10.1070/QE2010v040n02ABEH014206} {\bibfield
  {journal} {\bibinfo  {journal} {Quantum Electron.}\ }\textbf {\bibinfo
  {volume} {40}},\ \bibinfo {pages} {139} (\bibinfo {year} {2010})}\BibitemShut
  {NoStop}%
\end{thebibliography}
%

\appendix
\section{Explicit form of Eq.~(\ref{eq:chi33}) and Eq.~(\ref{eq:chi31})}

\begin{widetext}

\begin{equation}
 \langle \rho_{33} \rangle =
 \frac{\Omega_1^2 \Omega_2^2 \left(\Gamma_1 \Gamma_2 \left((\delta -\Delta)^2+(\Gamma_1+\Gamma_2)^2\right)+\Gamma_1 \Omega_1^2
   (\Gamma_1+\Gamma_2)+\Omega_2^2 (\Gamma_1+\Gamma_2)^2\right)}
   {K}
\end{equation}
\begin{equation}
\begin{split}
 | \langle \rho_{31} \rangle |^2 =&
 \left|\frac{\Omega_1 \Omega_2}
 {K}
 \right|^2\cdot\left|
 \delta ^3 \Gamma_1 \Gamma_2 (\Delta-i \Gamma_1)-\delta ^2 \Gamma_1 \Gamma_2 \left((\Delta-i
   \Gamma_1) (2 \Delta+i \Gamma_2)+\Omega_1^2+\Omega_2^2\right)\right.\\
 &\qquad\qquad\left.+\delta  \Gamma_1 \left(\Gamma_2 (\Delta-i \Gamma_1)
   \left(\Delta^2+2 i \Delta \Gamma_2+(\Gamma_1+\Gamma_2)^2\right)+\Omega_2^2 (\Delta (\Gamma_1+3 \Gamma_2)-i \Gamma_1
   (\Gamma_1+\Gamma_2))+2 i \Gamma_2 \Omega_1^2 (\Gamma_1+\Gamma_2)\right)\right.\\
&\qquad\qquad\left.-i \Delta^3 \Gamma_1 \Gamma_2^2-\Delta^2
   \Gamma_1 \Gamma_2 \left(\Gamma_1 \Gamma_2-\Omega_1^2+\Omega_2^2\right)-i \Delta \Gamma_1 \Gamma_2 (\Gamma_1+\Gamma_2)
   \left(\Gamma_2 (\Gamma_1+\Gamma_2)+2 \Omega_1^2+\Omega_2^2\right)\right.\\
&\qquad\qquad\left.-\left(\Gamma_1 \Gamma_2 (\Gamma_1+\Gamma_2)+\Gamma_1
   \Omega_2^2-\Gamma_2 \Omega_1^2\right) \left(\Gamma_1 \left(\Gamma_2 (\Gamma_1+\Gamma_2)+\Omega_1^2\right)+\Omega_2^2
   (\Gamma_1+\Gamma_2)\right)
   \right|^2
\end{split}
\end{equation}

with
\begin{equation}
\begin{split}
K=&
\delta ^4 \Gamma_1 \Gamma_2 \left(\Delta^2+\Gamma_1^2+2
   \Omega_1^2\right)-2 \delta ^3 \Delta \Gamma_1 \Gamma_2 \left(\Delta^2+\Gamma_1^2+2 \Omega_1^2+\Omega_2^2\right)\\
   &+\delta ^2
   \left(\Omega_2^2 \left(\Delta^2 \Gamma_1 (\Gamma_1+5 \Gamma_2)+\Gamma_1^2 \left(\Gamma_1^2+\Gamma_1 \Gamma_2+2
   \Gamma_2^2\right)+2 \Omega_1^2 (\Gamma_1+\Gamma_2)^2\right)\right.\\
&\quad\left.+\Gamma_1 \Gamma_2 \left(\Delta^2+\Gamma_1^2+2 \Omega_1^2\right)
   \left(\Delta^2+\Gamma_1^2+2 \Gamma_1 \Gamma_2+2 \Gamma_2^2-2 \Omega_1^2\right)+\Gamma_1 \Gamma_2 \Omega_2^4\right)\\
   &+2 \delta
   \Delta \left(-\Gamma_2 \Omega_2^2 \left(\Gamma_1 \left(\Delta^2+\Gamma_1^2+4 \Gamma_1
   \Gamma_2+\Gamma_2^2\right)+\Gamma_2 \Omega_1^2\right)+\Gamma_1 \Gamma_2 \left(\Omega_1^2-\Gamma_2^2\right)
   \left(\Delta^2+\Gamma_1^2+2 \Omega_1^2\right)-\Gamma_1 \Omega_2^4 (\Gamma_1+2 \Gamma_2)\right)\\&
   +\Delta^4 \Gamma_1\Gamma_2^3+\Delta^2 \Gamma_2 \left(\Gamma_1 \left(\Gamma_2^2 \left(2 \Gamma_1^2+2 \Gamma_1 \Gamma_2+\Gamma_2^2\right)+2
   \Gamma_2 \Omega_1^2 (\Gamma_1+2 \Gamma_2)+\Omega_1^4\right)+\Gamma_2 \Omega_2^2 \left(\Gamma_1 (3
   \Gamma_1+\Gamma_2)+\Omega_1^2\right)+\Gamma_1 \Omega_2^4\right)\\&
   +\left(\Gamma_2 (\Gamma_1+\Gamma_2)+\Omega_1^2+\Omega_2^2\right)
   \left(\Gamma_1^2 \Gamma_2+\Gamma_1 \Omega_2^2+2 \Gamma_2 \Omega_1^2\right) \left(\Gamma_1 \left(\Gamma_2
   (\Gamma_1+\Gamma_2)+\Omega_1^2\right)+\Omega_2^2 (\Gamma_1+\Gamma_2)\right)\,,
\end{split}
\end{equation}
\end{widetext}
where
$\Gamma_1$ and $\Gamma_1$ are the linewidths of the transitions addressed by pump~1 and~2, respectively.

\end{document}